\documentclass[aps,prl,twocolumn,showpacs,superscriptaddress,groupedaddress,notitlepage]{revtex4-1}
\usepackage[english]{babel}
\usepackage{amsmath,amssymb}
\usepackage{graphicx}
\usepackage{dcolumn}
\usepackage{bm}
\usepackage{amsmath}
\usepackage{xcolor}
\usepackage[applemac]{inputenc}
\usepackage{mathtools}

\usepackage{color}
\renewcommand{\vec}{\mathbf}

\renewcommand{\phi}{\varphi}

\newcommand{\type}[1]{\uppercase{S}^{#1}}
\newcommand{\vectype}[1]{\mathbf{\uppercase{S}}^{#1}}

\begin{document}


\hspace{5.2in} \mbox{Fermilab-Pub-04/xxx-E}

\title{Transitions in systems with High-Dimensional Stochastic Complex Dynamics: Monitoring and Forecasting }
\author{Duccio Piovani}
\email{duccio.piovani@gmail.com}
\author{Jelena Gruji\'c}
\email{jelenagr@gmail.com}
\altaffiliation{Current address: AI lab, Computer Science Department, Vrije Universiteit Brussel, Pleinlaan 2, 1050 Brussels, Belgium
and 
MLG, DÈpartement d'Informatique, UniversitÈ Libre de Bruxelles, Boulevard du Triomphe CP212, 1050 Brussels, Belgium}
\author{Henrik Jeldtoft Jensen}
\email{h.jensen@imperial.ac.uk}
\affiliation{Centre for Complexity Science and Department of Mathematics, Imperial College London, South Kensington Campus, SW7 2AZ, UK}
\begin{abstract}
We analyst in detail a new approach to the monitoring and forecasting of the onset of transitions in high dimensional complex systems (see Phys. Rev. Lett . {\bf 113}, 264102 (2014)) by application to the Tangled Nature Model of evolutionary ecology and high dimensional replicator systems with a stochastic element. A high dimensional stability matrix is derived for the mean field approximation to the stochastic dynamics. This allows us to determine the stability spectrum about the observed quasi-stable configurations. From overlap of the instantaneous configuration vector of the full stochastic system with the eigenvectors of the unstable directions of the deterministic mean field approximation we are able to construct a good early-warning indicator of the transitions occurring intermittently.  Inspired by these findings we are able to suggest an alternative  simplified applicable forecasting procedure which only makes use of observable data streams.
\end{abstract}

\maketitle

\textit{Introduction} - High dimensional complex systems both physical and biological exhibit intermittent dynamical evolution consisting of stretches of relatively little change interrupted by often sudden and dramatic transitions to a new meta-stable configuration~\cite{Sibani2013}. 

Such transitions can have crucial consequences when they occur in, say, ecosystems or financial markets and it is therefore important to develop methods that are able to identify precursors, warning signals and ideally techniques to forecast the transitions before they take place. We will expect that the mechanisms behind the rapid rearrangement may be different in different systems. Scheffer and collaborators have developed a method pertinent to systems in which the transition takes the form as a bifurcation captured by a robust macroscopic variable, which emerges from the micro dynamics. A precursor of the systemic change can then be identified from the critical slowing down and enhanced fluctuations exhibited by this macroscopic collective degree of freedom~\cite{scheffer2009critical,scheffer2012anticipating,scheffer2009early} as a change in some external parameter drives the system towards the bifurcation point. 

Here we consider an alternative scenario suggested recently in~\cite{Tran_Short} in which the transitions are induced by intrinsic fluctuations at the level of the individual components which propagates to the macroscopic systemic level and thereby triggers a change in the overall configuration. Our approach is relevant to systems in which the available configuration space evolves as a consequence of the dynamics. One may think of a new and more virulent virus being created through a mutation of an existing strain (e.g. the SARS virus in 2003), or a new economic agent arriving in the market (e.g. the dot-com bubble in 1997-2000). 

We describe below our methodology through applications to two models. First we consider the Tangled Nature (TaNa) Model of evolutionary ecology~\cite{tana:article1}, which has had considerable success in reproducing both macro-evolutionary aspects such as the intermittent mode of extinctions~\cite{tana:article2} and ecological aspects such as species abundance distributions~\cite{tana:article3} and species area laws~\cite{tana:article9}.

We also present results for transitions in a model with a very different type of dynamics, namely a high dimensional replicator with a stochastic element of mutation~\cite{diederich1989replicators,tokita2003emergence,PhysRevE.78.031924}. We demonstrate below that the replicator system with this element of stochasticity exhibit intermittency. Given the broad relevance of the replicator dynamics (population dynamics, game theory, financial dynamics, social dynamics etc.), success in forecasting transitions in this model may indicate that our method can be useful in many very different situations~\cite{kianercy2014critical}

Despite their different general mechanisms, the two models can be pictured in the same way. Their stochastic dynamics is characterized by a huge number of fixed points, and when the system randomly falls into one of them it enters a quiescent period of little change. Eventually the intrinsic stochastic fluctuations will allow the population of hitherto empty parts of configuration space, which may effectively serve as a random kick that is able to drive the system away from the local minimum and towards the chaotic regime where the system undergoes a high dimensional adaptive walk searching for another (metastable) fixed point.

Indeed both the nature of the fixed points and their stability varies significantly. Some fixed points are controlled by only a few interacting components while others involve many. Some are very stable while others less so leading to a very broad distribution of time spend in the metastable configurations of a given fixed point. The dynamics of the transitions between metastable configurations - the adaptive walk mentioned above - can also differ much. It can happen that the system is "trapped" between two or more attractors and switches between them before being pushed away. The transitions that lead from a fixed point to the other can be both sudden or slow and differ in magnitude. The point to be stressed is that the phenomenon we are trying to predict is highly heterogeneous and one has to bare this well in mind when interpreting the results.

That said, our claim is that we are able, in both models, to understand which kind of intrinsic stochastic fluctuation will be able to push the system out of its stable configuration. Indeed through a mean field description of the stochastic dynamics we can infer the Jacobian, from which by Linear Stability Analysis (LSA) we can identify the unstable eigendirections responsible for the destruction of the current metastable configuration.

As will be shown in the following sections, monitoring the relationship (vectorial overlap) between the existing configuration and the unstable mean field eigendirections dangerous directions allows to forecast approaching transitions with a high accuracy.

\section{Outline of forecasting procedure}
\label{OutLine}
In this section we first sketch our approach for then in the following two sections to describe in detail how to apply the method to the Tangled Nature Model and to the replicator system. 
 The first step is to establish a mean field approximation of the stochastic dynamics in order to obtain a set of deterministic equations. We establish the average flow of occupancy between different types of individual agents. Define the state vector $\vec{n}(t)=(n_1(t),\ldots,n_d(t))$, the mean field time evolution is of the form 
\begin{equation}
\vec{n}(t+1) - \vec{n}(t) = \mathbb{T}(\vec{n}(t))\vec{n}(t) 
\label{eq:mf2}
\end{equation}
where the matrix $\mathbb{T}$ is the mean field evolution matrix, which will contain contributions from the following processes: death, reproduction and mutation, and $\vec{n}(t)$ is a local time average of the stochastic configuration. 

\begin{figure}[h]
\centering

\includegraphics[width=8cm,height=5.34cm]{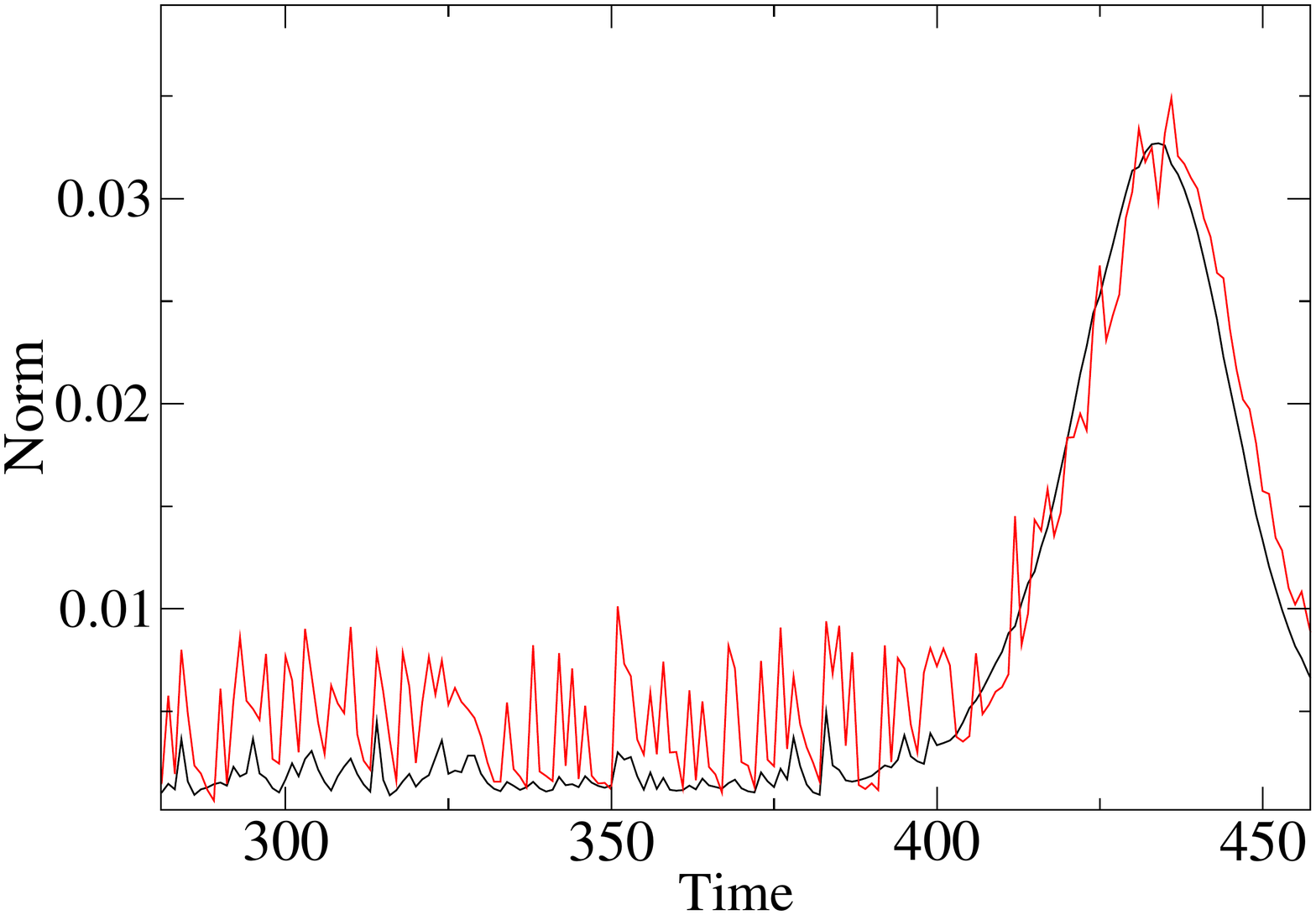}
\includegraphics[width=8cm,height=5.34cm]{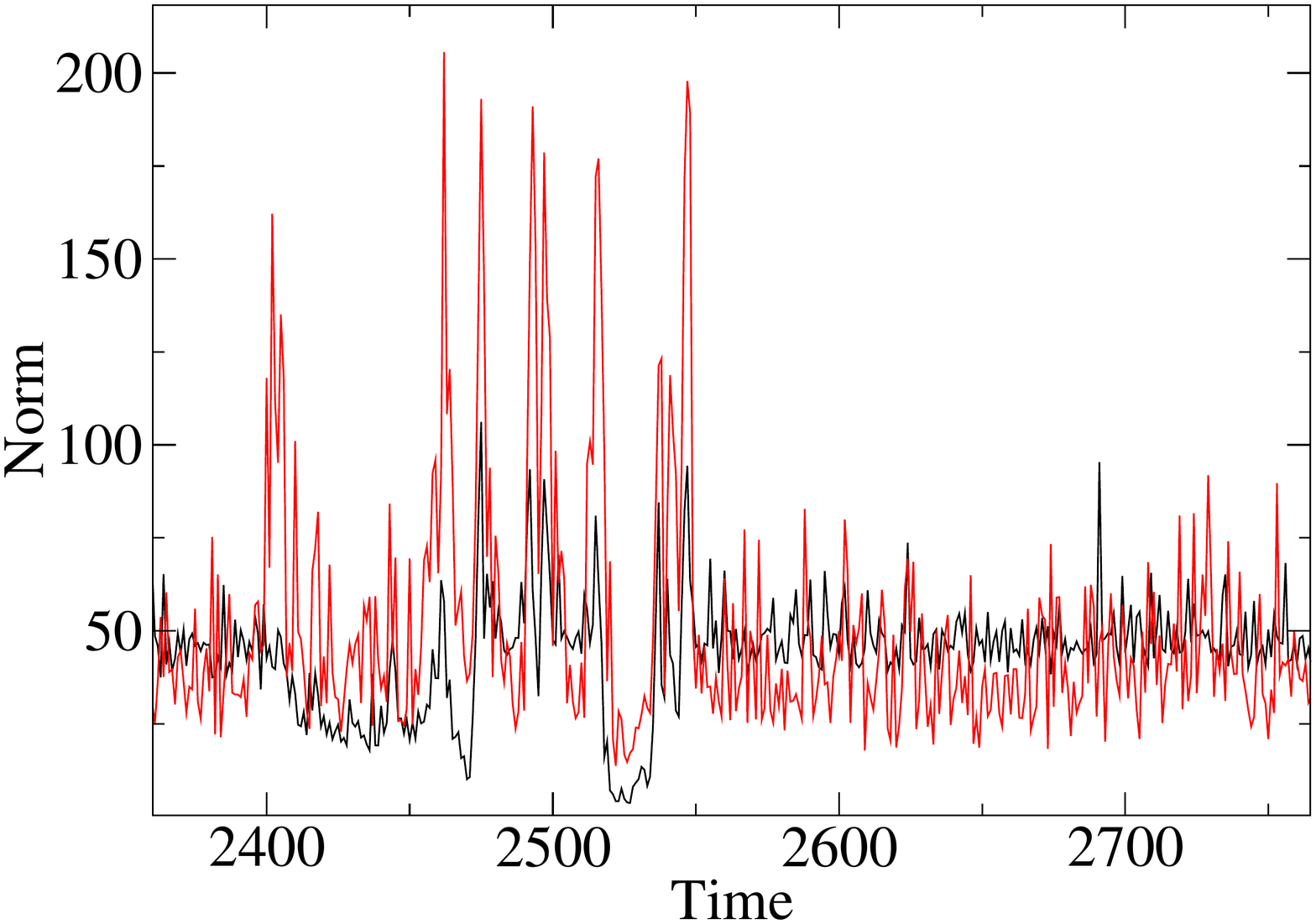}
\caption{ In this figure we compare the stochastic model (red curve) with its mean field approximation (black curve). At each time step the stochastic configuration is used as input in the mean field equation and the norms of the configuration vector are then compared. The top panel is for the Replicator Model while the Tangled Nature is in the right panel.}
\label{fig:mf_stoc}
\end{figure}

We can check the accuracy of our mean-field description of the stochastic system by measuring the norm of left hand side of Eq. (\ref{eq:mf2}), that is $\| \Delta n (t) \|$, during the simulations and compare it with the norm of the right hand side, i.e. $\| \mathbb{T}(\vec{n(t)})\cdot \vec{n}(t) \|$. If the difference
\begin{equation}
D({\rm sim},{\rm mean field})\equiv\| \Delta n (t) \|_\text{sim}- \| \mathbb{T}(\vec{n(t)}) \vec{n}(t) \|
\label{eq:check}
\end{equation}
is close to zero the mean field approximation will represent the stochastic dynamics well, at least in a local time and configuration neighbourhood of $\vec{n}(t)$. This suggests then that we can use Eq.(\ref{eq:mf2}) to study local stability properties. In Fig. (\ref{fig:mf_stoc}) we can see how these 2 quantities relate in the 2 models. In the Replicator Model (left panel) the mean field evolution (black curve) appears to be the average of the stochastic evolution (red curve), while in the Tangled Nature Model (right panel), they differ more clearly. This result depends on the different type of dynamics of the models, the Tangled Nature being completely stochastic while the Replicator Model being more close to a Langevin dynamics. 
 
Obviously in the mean field approximation the fixed point configurations are given as solutions to $\mathbb{T}(\vec{n}(t))\cdot \vec{n}(t) $, see Eq. (\ref{eq:mf2}). Because of the high dimensionality of the type of systems we have in mind, this equation will typically not be solvable analytically. In any case, the stochastic dynamics will not satisfy the fixed point conditions strictly. Rather we'll expect little time variation during a meta stable phase, i.e. $\vec{n}(t+1) \simeq \vec{n}(t) \simeq \vec{n}^*$, where $\vec{n}^*$ is a local time average of $\vec{n}(t)$. This means that the left hand side of Eq. (\ref{eq:mf2}) will be close to zero and that $\vec{n}^*$ is essentially a fixed point of the mean field dynamics. We perform a linear stability analysis about $\vec{n}^*$ by expanding the right hand side of Eq. (\ref{eq:mf2}). We introduce $\vec{n}(t) = \vec{n}^* + \delta \vec{n}(t)$, expand to first order in $\delta\vec{n}(t)$ get from Eq. (\ref{eq:mf2})
\begin{equation}
\delta \vec{n}_i(t+1) - \delta \vec{n}_i(t)  \simeq  \left( \mathbb{T}(\vec{n}^*) + \partial_\vec{n}\mathbb{T} \vec{n}^* \right)\delta \vec{n}
\end{equation}
\[
= \mathbb{M}(\vec{n}^*) \delta \vec{n}^*
\]
here the matrix 
\begin{equation}
\mathbb{M}(\vec{n}^*) =  \left( \mathbb{T}(\vec{n}^*) + \partial_{n} \mathbb{T}(\vec{n}^*)\vec{n}^* \right) 
\label{eq:M}
\end{equation}
is the \emph{Jacobian} of the system, or the \emph{stability matrix}. Now exploiting the results of the LSA, we know that the eigenvectors or \emph{generalised} eigenvectors (in case of a non diagonalizable Jacobian) $\vec{e}_+ $ associated with $\lambda$ with $\text{Re}(\lambda)>0$ indicate unstable directions. These can be identified with \emph{toxic} components $n_\text{t}$ of the configuration vector. 

What this means is that if the stochastic fluctuations bring the system towards these unstable directions, by activating the \emph{toxic} components, the system would feel a repulsive force that would push it away from $\vec{n}^*$. 
A sudden growth of these components would indicate the arrival of a transition. This observation allows us to identify a stability indicator,who's non-zero values are early warning signaling of an approaching transition caused by the system leaving the vicinity of a current fixed point. The details of this indicators will depend on the specific case we are dealing with but will be based on the same general idea.

\onecolumngrid

\noindent\makebox[\linewidth]{\rule{\textwidth}{1pt}} 
\vspace{.5cm}
\twocolumngrid

In the following sections we will present the two model's, analyzing their basic mechanisms, and develop our mean-field stability indicator in both cases. 

\section{The Tangled Nature Model} 
 
\begin{figure}[h]
\centering
\includegraphics[width=8cm,height=5.34cm]{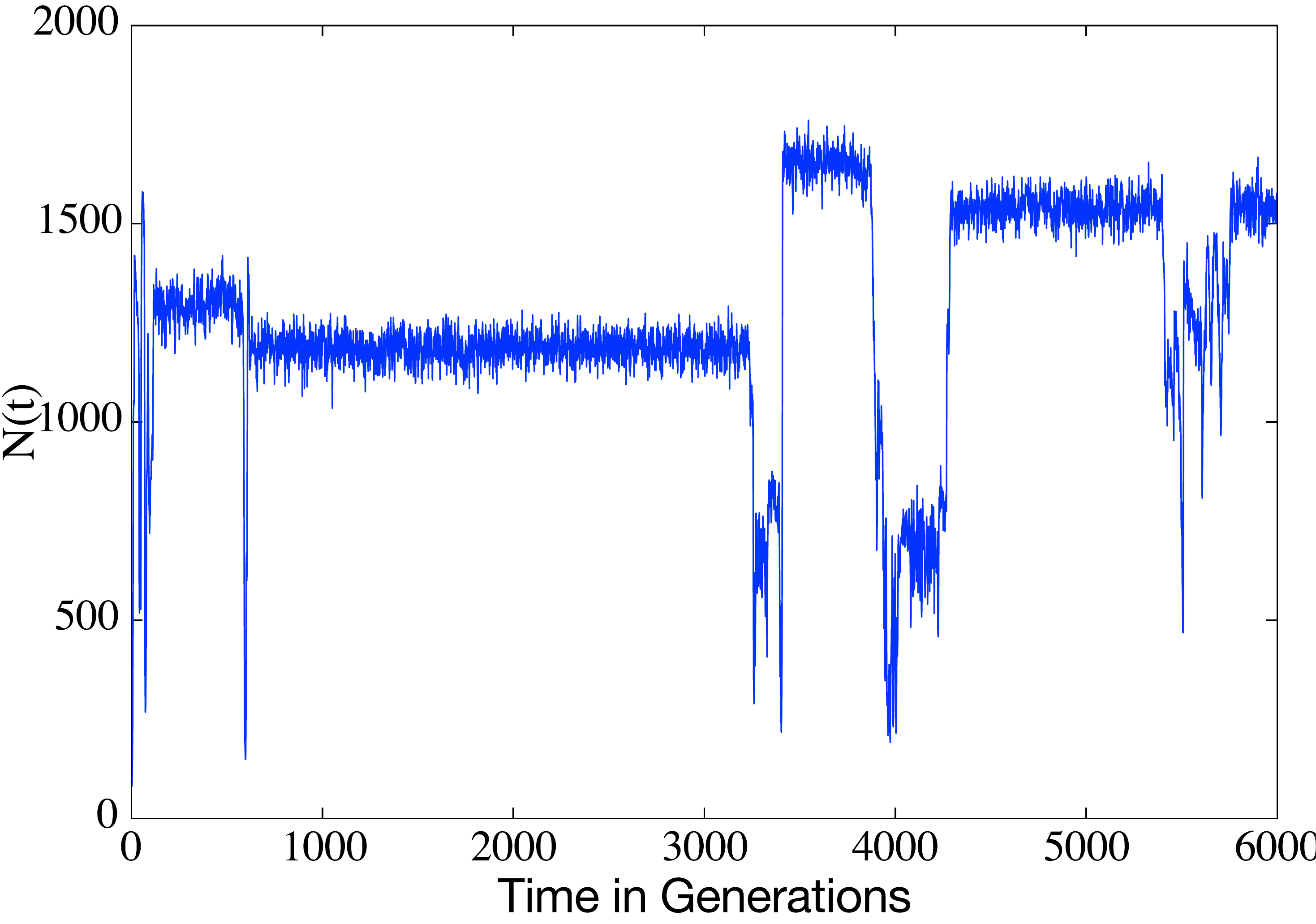}
\includegraphics[width=8cm,height=5.34cm]{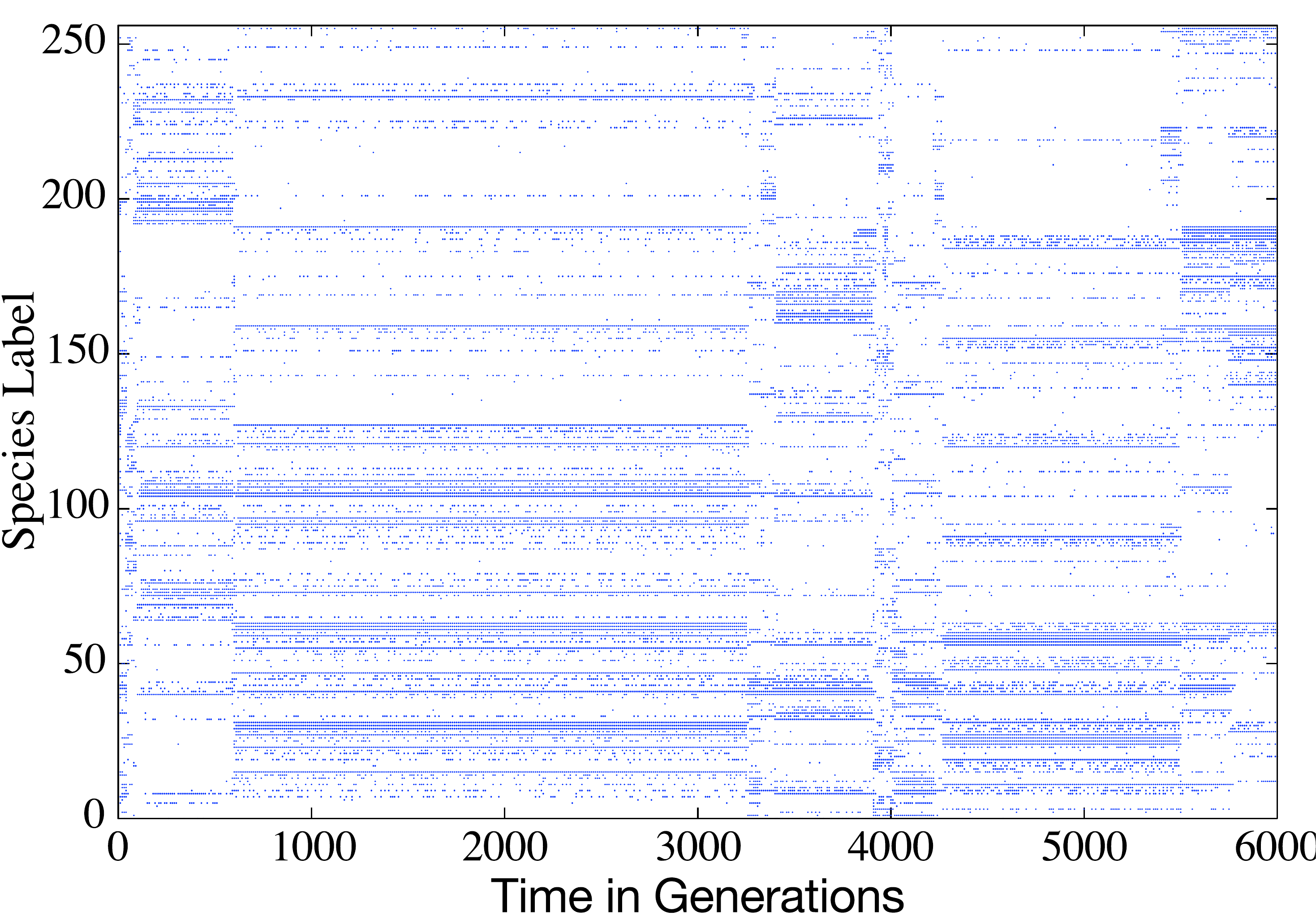}
\caption{Left panel: total population as a function of time (in generations) for a single realization of the TaNa. The punctuated dynamics is clearly visible: quasi-stable periods alternate with periods of hectic transitions, during which $N(t)$ exhibits large amplitude fluctuations. Right panel: occupancy distribution of the types. The genotypes are labelled arbitrarily and a dot indicates a type which is occupied at the time t. These figures are obtained with parameters $L=8$, $p^\text{mut} =0.2$, $p^\text{kill}=0.4$, $K=40$ and $\mu = 0.07$.}
\label{fig:1}
\end{figure}

 \textit{The model} - In the TaNa, an agent is represented by a sequence of binary variables with fixed length L~\cite{Higgs/Derrida:article}, denoted as $\vectype{a}=(\type{a}_1,\ldots,\type{a}_L)$, where $\type{a}_{i}=\pm 1$. Thus, there are $2^L$ different sequences, each one represented by a vector in the genotype space: $\mathcal{S}=\{-1,1\}^{L}$. In a simplistic picture, each of these sequences represents a genome uniquely determining the phenotype of all individuals of this genotype. We denote by $n({\bf S}^a,t)$ the number of individuals of type ${\bf S}^a$ at time $t$ and the total population is $N(t)=\sum_{a=1}^{2^L} n(\vectype{a},t)$. We define the distance between different genomes $\vectype{a}$ and $\vectype{b}$ as the Hamming distance: $d_{ab}=\frac{1}{2L}\sum_{i=1}^{L}|\type{a}_{i}-\type{b}_{i}|$. A time step is defined as a succession of one annihilation and of one reproduction attempt. During the killing attempt, an individual is chosen randomly from the population and killed with a probability $p_{kill}$ constant in time and independent of the type. During the reproduction process, a different randomly chosen individual $\vectype{a}$ successfully reproduces with probability: $p_{off}(\vectype{a},t)=\frac{\exp{(\uppercase{h}(\vectype{a},t))}}{1+\exp{(\uppercase{h}(\vectype{a},t))}}\label{eq:2.3}$, which depends on the occupancy distribution of all the types at time $t$ via the weight function:
 \begin{equation}
\uppercase{h}(\vectype{a},t)=\frac{k}{\,\uppercase{n}(t)}\sum_{\vectype{b}\in\,\mathcal{S}}\mathbf{J}(\vectype{a},\vectype{b})n(\vectype{b},t)-\mu\uppercase{n}(t).
\label{eq:1}
\end{equation}   
In Eq. (\ref{eq:1}), the first term couples the agent $\vectype{a}$ to one of type $\vectype{b}$ by introducing the interaction strength $\mathbf{J}(\vectype{a},\vectype{b})$, whose values are randomly distributed in the interval $(-1,+1)$. For simplicity and to emphasize interactions we here assume: $\mathbf{J}(\vectype{a},\vectype{a})=0$. The parameter $k$ scales the interactions strength and $\mu$ can be thought of as the carrying capacity of the environment. An increase (decrease) in $\mu$ corresponds to harsher (more favorable) external conditions. The reproduction is asexual: the reproducing agent is removed from the population and substituted by two copies $\vectype{a}_1$ and $\vectype{ a}_2$, which are subject to mutations. A single mutation changes the sign of one of the genes: $\type{\gamma}_{i}\rightarrow -\type{\gamma}_{i}$ with probability $p_{mut}$. Similarly to a Monte Carlo sweep in statistical mechanics, the unit of time of our simulations is a \emph{generation} consisting of $N(t)/p_{kill}$ time steps, i.e. the average time needed to kill all the individuals at time $t$. These microscopic rules generate intermittent macro dynamics. The system is persistently switching between two different modes: the meta-stable states (denoted quasi-Evolutionary Stable Strategies or qESS) and the transitions separating them. The qESS states are characterized by small amplitude fluctuations of N(t) and stable patterns of occupancies of the types (Fig. \ref{fig:1}, respectively top and bottom panel). However, these states are not perfectly stable and configurational fluctuations may trigger an abrupt transition to a different qESS state. The transitions consist of collective adaptive random walks in the configuration space while searching for a new metastable configuration and are related to high amplitude fluctuations of N(t). All the results we will present for this model have been obtained fixing the parameters to $L=8$, $p^\text{mut} =0.2$, $p^\text{kill}=0.4$, $K=40$ and $\mu = 0.07$.
 
\onecolumngrid

\vspace{2cm}

\twocolumngrid
 
\subsection{Mean Field Description - Tangled Nature Model}
In the TaNa model there are multiple sources of stochasticity, namely reproduction, mutations and deaths. Following the procedure outlined above we average out these sources and formulate a deterministic mean field equation. 
At each time step with probability $p_\text{kill}$ a randomly chosen individual is removed from the system, which implies that the occupation number of the species it belongs to decreases of 1 unit ($\Delta n_i = -1$). Given that the probability of choosing an individual belonging to the $ith$ species is $\rho_i = \frac{n_i}{N}$, the killing term becomes
\begin{equation}
\rho_i   p_\text{kill}   (-1).
\label{eq:kill}
\end{equation}
The reproduction term is slightly more complicated given the presence of mutations. A randomly chosen individual is selected for asexual reproduction, which means it is removed from the system while creating two new individuals of the same species. Offsprings can both mutate ($\Delta n_i  = -1 $ ), only one can mutate ($\Delta n = 0 $), or none mutate ($\Delta n = +1$). Keeping in mind that the probability of reproducing is given by $p^\text{off}_i$  the average contribution from mutation is 
\begin{equation}
 \rho_i    p_i^\text{off}(t)\left[ 2p_o - 1 \right] = \alpha    \rho_i    p_i^\text{off}(t) 
\end{equation}
here $p_o =(1-p^\text{mut})^L$ is the probability of no mutations and $\alpha  = (2p_o -1)$ is a constant. The third term we have to consider is the \emph{back-flow effect}, which describes the event of begin populated by mutations occurring during the reproduction happening elsewhere. This term has the form 
\begin{equation}
\sum_{j} \rho_{j}(t) p_j^\text{off} p^\text{mut}_{j \rightarrow i}.
\end{equation}
For a type Hamming distance $d_{ij}$ away to be able to mutate on to a given type $d_{ij}$ genes will have to mutate and this will happen with the binomial probability
\begin{equation}
p^\text{mut}_{i \rightarrow j} =p_\text{mut}^{d_{ij}}   (1-p_\text{mut})^{L-d_{ij}}.
\end{equation}
Putting together all these effects we find the form of Eq.(\ref{eq:mf2}) for this model, namely
\begin{equation}
n_i(t+1) - n_i(t) = \frac{1}{N} \sum_j \{  \left(   p_j^\text{off}(t)  \left( 2p_o - 1 \right) - p^\text{kill}\right)   \delta_{ij} + 
\end{equation}
\[
 p_j^\text{off}   p^\text{mut}_{j \rightarrow i}  \left(1-\delta_{ij}\right)\} n_j(t)
\]
where 
\begin{equation}
T_{ij} =   \left(   p_j^\text{off}(t)  \left( 2p_o - 1 \right) - p^\text{kill}\right)   \delta_{ij} + p_j^\text{off}   p^\text{mut}_{j \rightarrow i}  \left(1-\delta_{ij}\right)
\label{mf:T}
\end{equation}
it is mean-field evolution matrix of the system.

By substituting Eq. (\ref{mf:T}) into Eq. (\ref{eq:M}) we get the specific form of the stability matrix for the Tangled Nature Model
\begin{equation}
\mathbb{M}_{ij} = (\alpha p^\text{off}_j - p^\text{kill})\delta_{ij} + 2(1-\delta_{ij})p_j^\text{off}p_{j\rightarrow i}^\text{mut}
\label{eq:Mtana}
\end{equation}
\[
+ \sum_k\left[ {\alpha\delta_{ik} + (1-\delta_{ik} )   p^\text{mut}_{k \rightarrow i}  }\right]\frac{\partial p_\text{k}^\text{off}}{\partial n_j} n^*_k.
\]
This is the mean field matrix we use for our linear stability analysis of the stochastic fixed points. 

\section{The Replicator Model with Stochasticity}
The deterministic version of the replicator dynamics~\cite{taylor1978evolutionary} is used routinely in a large variety of applications, not least because of its relation to game theory~\cite{kianercy2014critical, hofbauer1998evolutionary} and is therefore expected to be of relevance to the description of e.g. high dimensional socio-economic or biological systems. This suggests that if our method works for the stochastic replicator the procedure can be of broad relevance as a way to identify and analyse precursors of endogenous transitions.   

We are interested in the limit of many strategies played by agents that may leave the system (say go bankrupt or extinct) or may change their strategy, or mutate. This version of the replicator dynamics set-up was studied by Tokita and Yasutomi in~\cite{tokita2003emergence}. The authors focused on the emerging network properties. Here we continue this study but with an emphasis on the intermittent nature of the macro-dynamics.

For this model the configuration vector $\vec{n}$ contains the relative frequencies of all the allowed $d$ different frequencies, so the components $n_i(t)\in[0,1]$ for all $i=1,2,...,d$ but not all frequencies may be occupied at a given moment, i.e. we can have $n_i(t)=0$ for some strategy $i$. We start the simulations by generating the $d\times d$ payoff matrix $J$ of the game that will tells us the payoffs of every pairwise combination. As for the Tangled Nature Model above, the matrix $J$ is a random and constant interaction network on which the replicator dynamics will be embedded. Each strategy distinguishes itself from the others in its payoffs or interactions with the rest of the strategy space. 

In this chapter we used the same type of uncorrelated interaction matrix as used in the study above of Tangled Nature Model. The dimension of the matrix is large, namely $d\in ( 10^2 ,10^4)$. The qualitative aspects of the behavior remains the same for other types of payoff matrices. We found that matrices with payoffs uniformly distribute on the interval $(-1,1)$ or on the set $\{0,1\}$ exhibit the same behavior as matrix of the form used for the Tangled Nature Model. However, if the payoffs are drawn from a power law distribution with no second moment, the dynamics becomes different and the intermittent behaviors is not so clear any more.
  
In the initial configuration, $N_\text{o}\ll d$ strategies start with the same frequency $n_i = \frac{1}{N_\text{o}}$. All the other possible strategies are non active, i.e. the corresponding components $d-N_\text{o}$ in $\vec{n}(0)$ are $n_i(0)=0$, since no fraction of the players use them. The empty strategies can only become populated by one of the \emph{active} strategies mutating into them. Once this happens their frequency will evolve according to the replicator equation in which these newly occupied strategies interact with the active strategies which they are linked to through the matrix $J$. 

A time step of the replicator dynamics consists in calculating the \emph{fitness}, $h_i(t) =  \sum_j J_{ij}n_j(t) $ of each active strategy and compare it with the average fitness $\bar{h}(t)= \sum_{ij}J_{ij}n_{i}(t)n_{j}(t) $, exactly as expected in a replicator dynamics. Each frequency is then updated according to
\begin{equation}
n_{i}(t+1)= n_i(t)+\left( \sum_j J_{ij} n_j(t) - \sum_{ij}J_{ij}n_{i}(t)n_{j}(t) \right)  n_i(t)
\end{equation}
The stochastic element, of the otherwise deterministic dynamics, consists in the following updates. With probability $p^{\text{mut}}$ each strategy mutates into another one, this is done by transferring a fraction $\alpha_{\text{mut}}$ of the frequency from the considered strategy to another strategy. The label of the latter strategy is chosen in the vicinity of the first by use of a normal distribution $N(i,\Delta)$ centered on label $i$ with variance $\Delta$. The closer the labels of two strategies the more likely it is for one to mutate into the other.

It should be noted that as long as the payoff matrix is random and uncorrelated in its indices, no similarity criteria between strategies doesn't really exists ( 2 similar strategies interact in a completely different way with the environment ). The parameter has been introduced only to control the level of disorder of the system. 

When the frequency of a strategy $i$ goes below a preset extinction threshold $n_i(t)<n^{\text{ext}}$, the strategy is considered extinct and its frequency is set to zero $n_i(t+1) = 0$. Right after an extinction event the system is immediately renormalized in order to maintain the condition $\sum_i n_i(t) =1$.

\begin{figure}[!h]
\centering
\includegraphics[width=8cm,height=5.34cm]{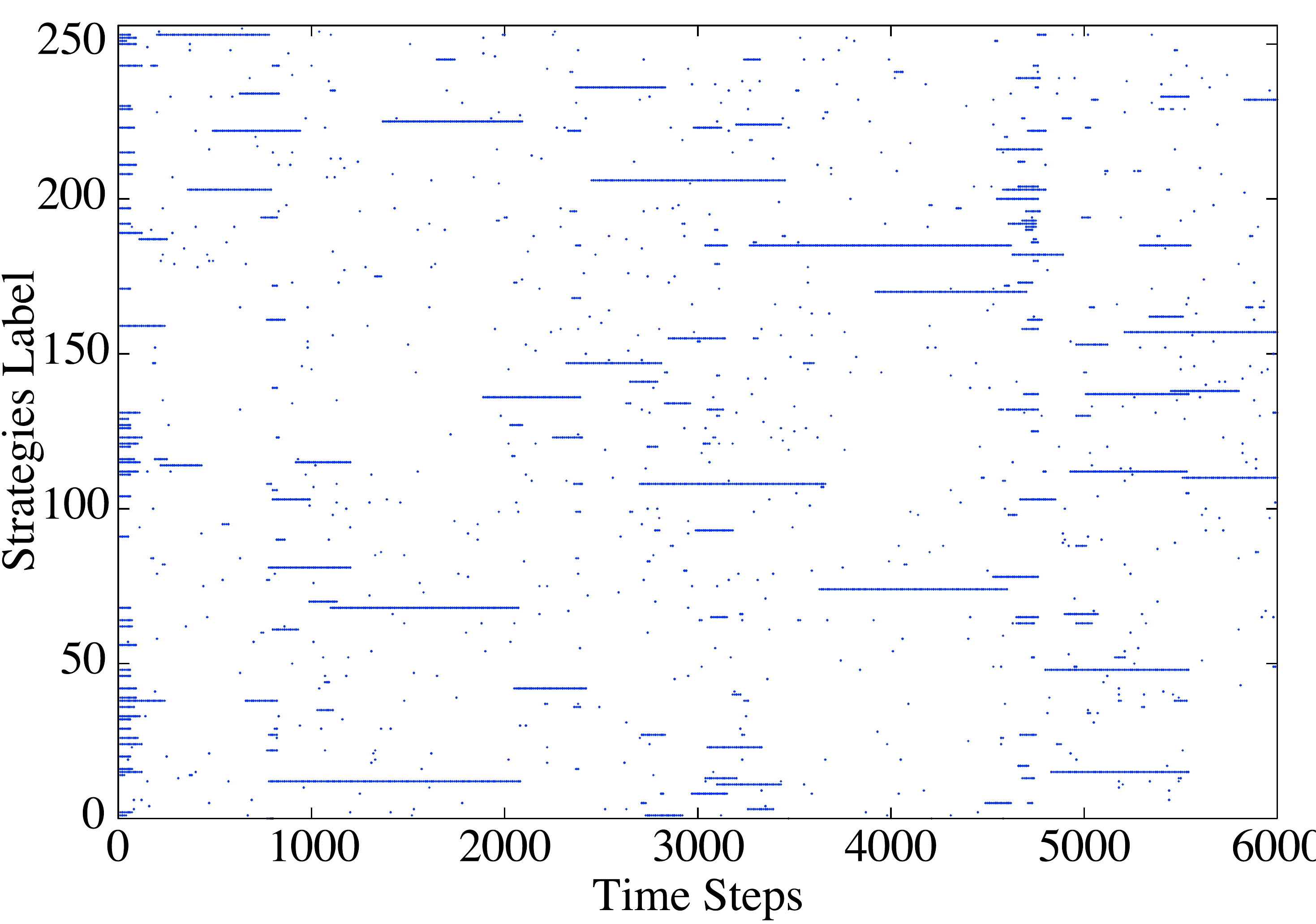}
\includegraphics[width=8cm,height=5.34cm]{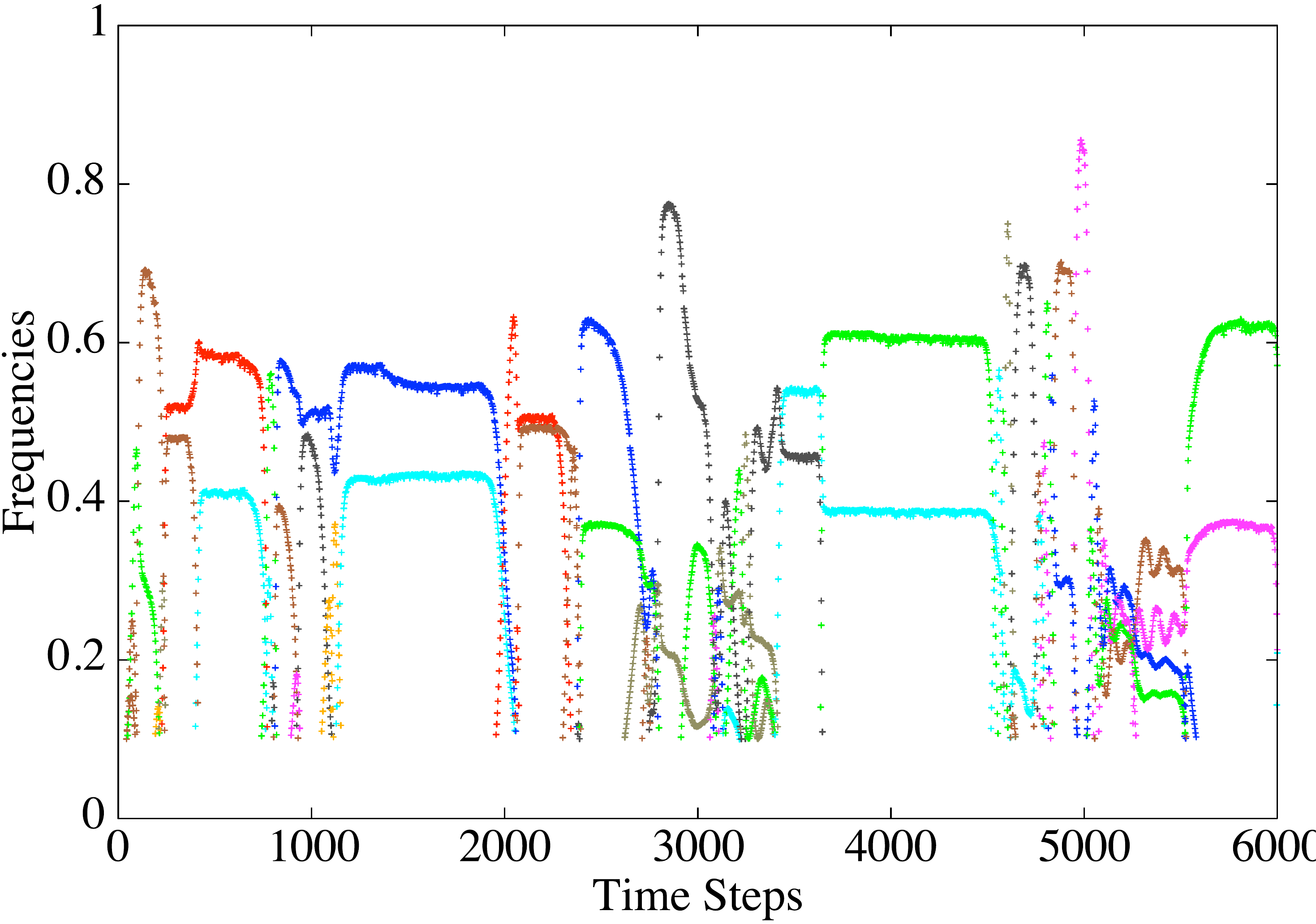}
\caption{Left panel: occupancy distribution of the types. The genotypes are labelled arbitrarily and a dot indicates a type which is occupied at the time t. The punctuated dynamics is clearly visible: quasi-stable periods alternate with periods of hectic transitions. Right panel: We present the frequencies of the strategies. Each color belongs to a different strategy. Once again the transitions from fixed point to another is clear.}
\label{fig:2}
\end{figure}

The systemic level dynamics exhibit complex dynamics as seen from the time evolution of the occupancy vector $\vec{n}(t)$, see in Fig.(\ref{fig:2}). 

All the results for this model have been obtained with the same parameter set, namely: $d = 256$, $n_\text{ext}=0.001$, $\alpha_\text{mut}=0.01$, $p^\text{mut} =0.2$.

\onecolumngrid

\noindent\makebox[\linewidth]{\rule{\textwidth}{1pt}} 
\vspace{0.5cm}
\twocolumngrid

\subsection{Mean Field Description - Replicator Model}

The random mutations are the only source of stochasticity in the model's dynamics. To account for these stochastic events one has to take into account the possibility that a strategy looses part of its frequency by mutating into other strategies and gaining frequency as a result of mutations happening elsewhere. As a result a given strategy may loose a fraction of players $\alpha_\text{mut}$, which happens with probability $p_\text{mut}$ or gain $\alpha_{\text{mut}} n_j(t+1)$ which happens with probability $\frac{p_\text{mut}}{N_a}$ where $N_a$ is the number of active strategies. This second effect describes the probability of being populated by a mutation. We therefore get the mean field description as 
\[
n_{i}(t+1)\simeq n_i(t)+\left( \sum_j J_{ij}n_j(t) - \sum_{ij}J_{ij}n_{i}(t)n_{j}(t) \right)  n_i(t)
\]
\begin{equation}
 -\alpha_{\text{mut}}p_\text{mut}  n_i(t) + \frac{p_\text{mut}}{N_a(t) }\sum_j \alpha_{\text{mut}}n_j(t) 
\label{eq:mf}
\end{equation}
which can be expressed, in compact form as
\begin{equation}
n_{i}(t+1)-n_i(t)\simeq \sum_j \mathbb{T}_{\text{ij}}n_j(t)
\label{eq:mfc}
\end{equation}
where
\begin{equation}
\mathbb{T}_{ij} = \left(  \sum_j J_{ij}n_j(t) - \sum_{ij}J_{ij}n_{i}(t)n_{j}(t) - \alpha_{\text{mut}}\right) \delta_{ij} 
\label{eq:T}
\end{equation}
\[
-\frac{\alpha_{\text{mut}}}{2^L-1} \left(1-\delta_{ij}\right)
\]
The stability matrix is obtained by substitution in eq.(\ref{eq:M}) 
\begin{equation}
\mathbb{M}_{ij} = \mathbb{T}_{ij}(\vec{n^*}) + \left[J_{ij}  - \sum_k(J_{ik} + J_{ki})\vec{n}_k^*\right]\vec{n}^*_i
\label{eq:Mrm}
\end{equation}

\section{Forecasting Procedure and Results}

We described in the previous sections how the dynamics in the two models consists in intermittent swift transitions between quasi-metastable configurations. As mentioned in the previous sections we approximate the fixed points of the mean field dynamics by local time averages over successive configurations in the quasi-stable phases of the full stochastic dynamics, namely: $\vec{n}^* =\frac{1}{T} \sum_{t=0}^T \vec{n}(t)$, which we will treat as our fixed point. 

Through our procedure we want to study the stability in the neighborhood of $\vec{n}^* $, in order to predict the system's reaction to the stochastic perturbations. To the extent that the mean field matrix correctly describes the system the metastable states will become unstable along directions in configuration space given by the eigenvectors $\vec{e}_+$ corresponding to eigenvalues with a positive real part $\text{Re}(\lambda)>0$.

Once we know the form of the eigenspace we can monitor two important scalar quantities: the instantaneous distance from the fixed point 
\begin{equation}
\delta n(t) = \| \delta \vec{n}(t) \| = \|\vec{n}(t) - \vec{n}^*\|  
\label{eq:delta}
\end{equation}  
and the maximum overlap between the perturbation and the eigenvectors $\{e_+\}$ of the unstable subspace
\begin{equation}
Q(t) = |\delta \vec{n}(t)  \vec{e}_i |_\text{max}
\label{eq:Q}
\end{equation}

\begin{figure}
\centering
\includegraphics[scale=0.33]{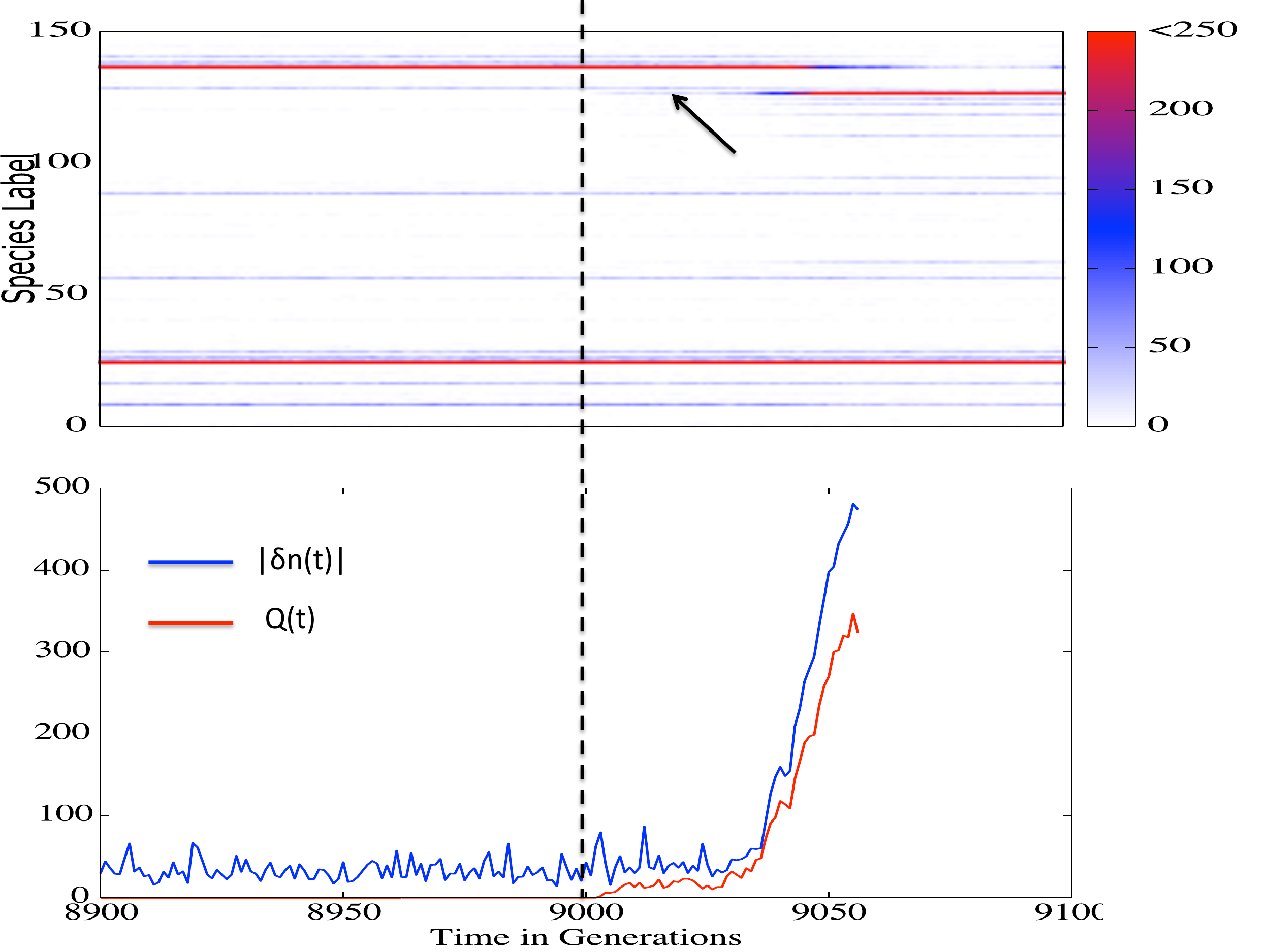}
\caption{In the bottom panel of this figure we show the behavior of $\delta n(t) $ (blue curve) and $Q(t)$ (red curve) while approaching the transition in the Tangled Nature. In the top panel a weighted occupation plot is presented. We can see how the beginning of the transitions (dashed vertical black line) is triggered by a new mutant (black arrow) that quickly gains population. The arrival of the new dangerous mutant is singled by a peak in the Q(t).}
\label{fig:transition_tana}
\end{figure}

The quantity in eq.(\ref{eq:delta}) tells us how far away the system is from the fixed point. If no unstable directions exist the system will be expected to stays in the vicinity of the fixed point and hence we expect $\delta n(t)$ to fluctuate around a low constant value, while a transition would induce a sudden increase in $\delta n(t)$. The overlap in eq.(\ref{eq:Q}) tells us to what extent a deviation $\vec{n}(t) - \vec{n}^*$ is within an unstable sub space.

Another way of picturing $Q(t)$ is as a measure of the activity of the occupancy on dangerous nodes. Indeed every non zero component of the unstable eigenvectors $\{\vec{e}_+\}$ will tell us which nodes of the interaction network are capable of pushing the system out of its metastable configuration. Namely if $e^j_+>0$, where $j$ indicates the component of the  unstable eigenvector, this means the the $jth$ node is dangerous.

\begin{figure}
\centering
\includegraphics[scale=0.33]{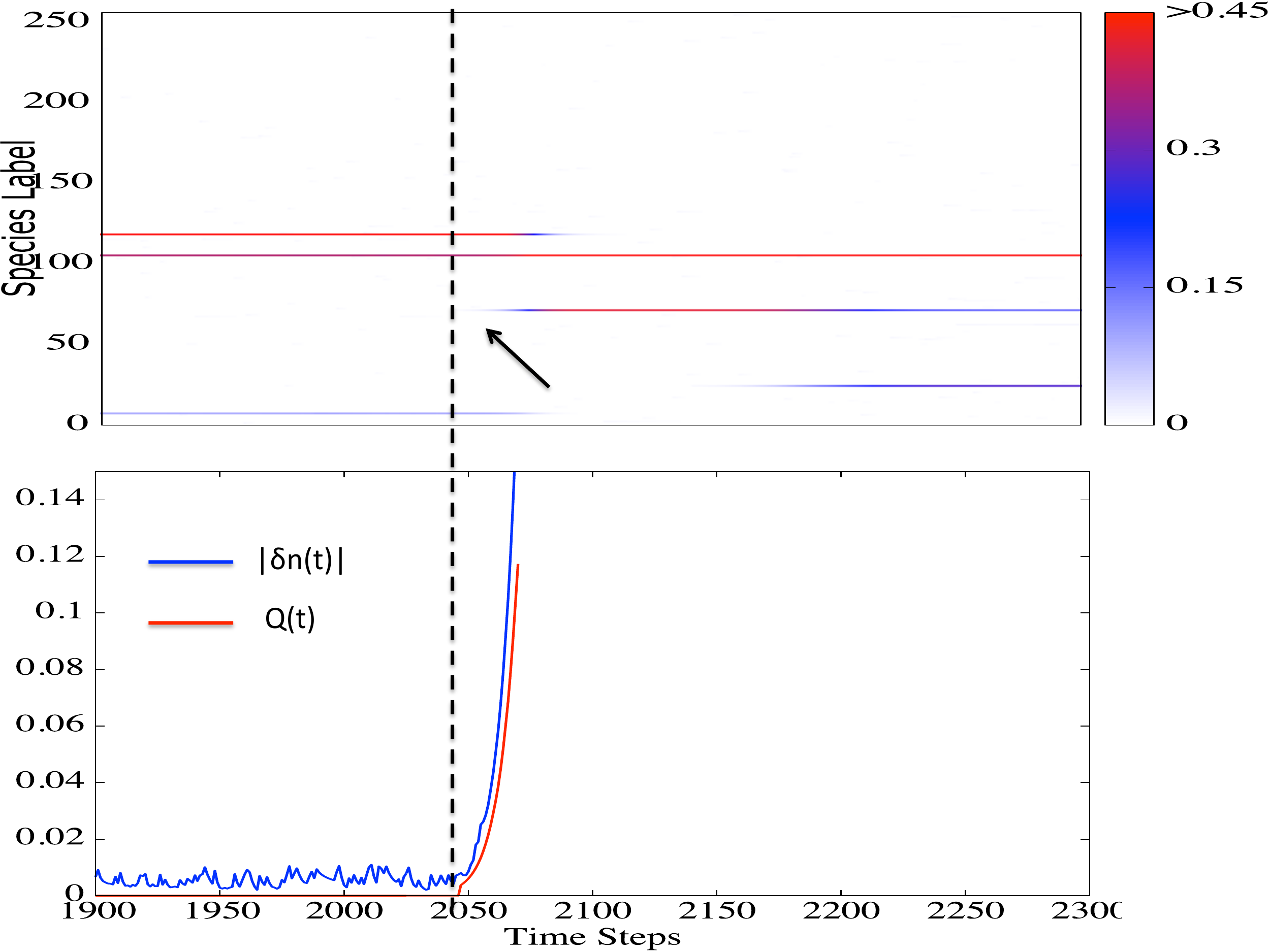}
\caption{ This is the same type of figure showed in Fig.(\ref{fig:transition_tana}) for the Replicator Model. Bottom panel $\delta n(t) $ and $Q(t)$, blue and red curve respectively, top panel weighted occupation plot. We can see how even in this model the transition is triggered but the arrival of a new fit mutant that my gaining weight disturbs the existing equilibrium.}
\label{fig:transition_rep}
\end{figure}

The $Q(t)$ monitor the activity of such nodes. If one of these nodes were to become activated by mutations (which the stochastic perturbations) this would result into a rapid growth of $Q(t)$ and can be considered as a warning of successive transition. 

 In~\cite{Tran_Short} in Fig.(3) it was discussed how these two quantities behave in the TaNa model and we demonstrated the forecasting power of the indicator $Q(t)$ and we gave an explanation on why we missed some of the transitions.
 
Here we illustrate in Fig.(\ref{fig:transition_rep}) and Fig.(\ref{fig:transition_tana}) the temporal behaviour of $Q(t)$ and $\delta n(t)$ for both the Tangled Nature Model and the stochastic replicator system. 
The top panels contain weighted occupation plots while the bottom figures show the behavior the two quantities in $Q(t)$ and $\delta n(t)$. The arrow points at the new dangerous mutant that has entered the system, while the dashed bar indicates the moment it happens. Before the dashed line we can see how fluctuations in $\delta n(t)$ are bounded and $Q(t)$ essentially equals to zero. After the dashed line, when the new mutant has entered the system, we see an explosion of both quantities. 
 
We denote $t^*$ the time at which the transition begins, which is set by the $\delta n(t)$ crossing a reasonably chosen threshold $T_\delta$ and staying consistently above this threshold (we have used $T_\delta=150$ for the TaNa and $T_\delta=0.05 $ for the Replicator Model). Given the sharp increase of $\delta n(t)$ when approaching the transition, $t^*$ doesn't depend strongly on the precise choice of the threshold as long as its is chosen larger than the characteristic fluctuations of $\delta n(t)$ during the metastable configurations. .

To define an alarm we determine an appropriate threshold $A_Q$ on $Q(t)$. To do so we compare the number of false alarms with the number of missed transitions generated by different values of the chosen threshold $A_Q$. We define a false alarm when the $Q(t)$ crosses $A_Q$ but then goes back under it before any transition occurs. On the other hand a missed transition corresponds to situations where $Q(t)$ remained below $A_Q$ even though the given metastable configuration did become unstable and therefore a transition did occur. 
 
\begin{figure}[ht!]
\centering
\includegraphics[width=8cm,height=5.34cm]{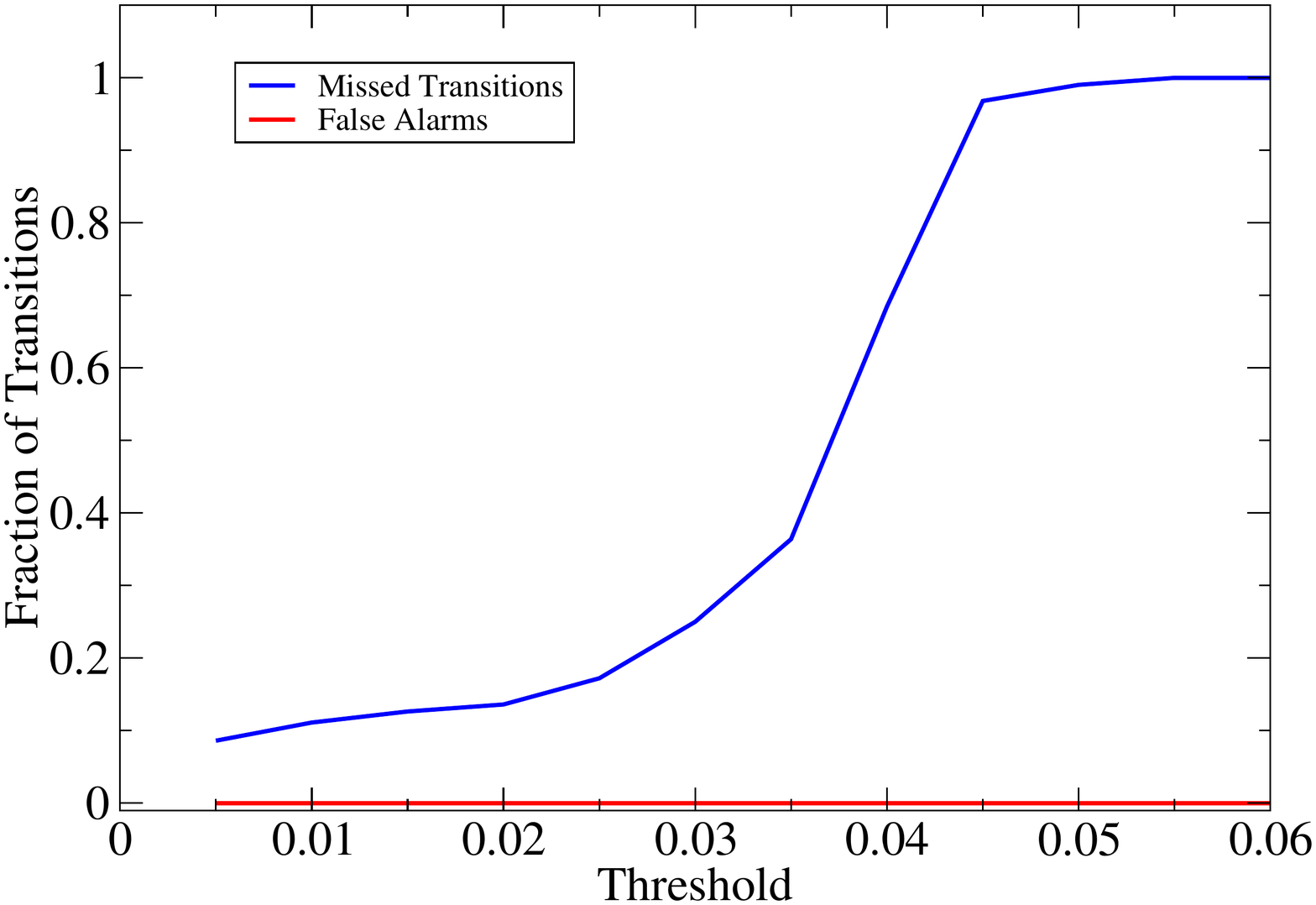}
\includegraphics[width=8cm,height=5.34cm]{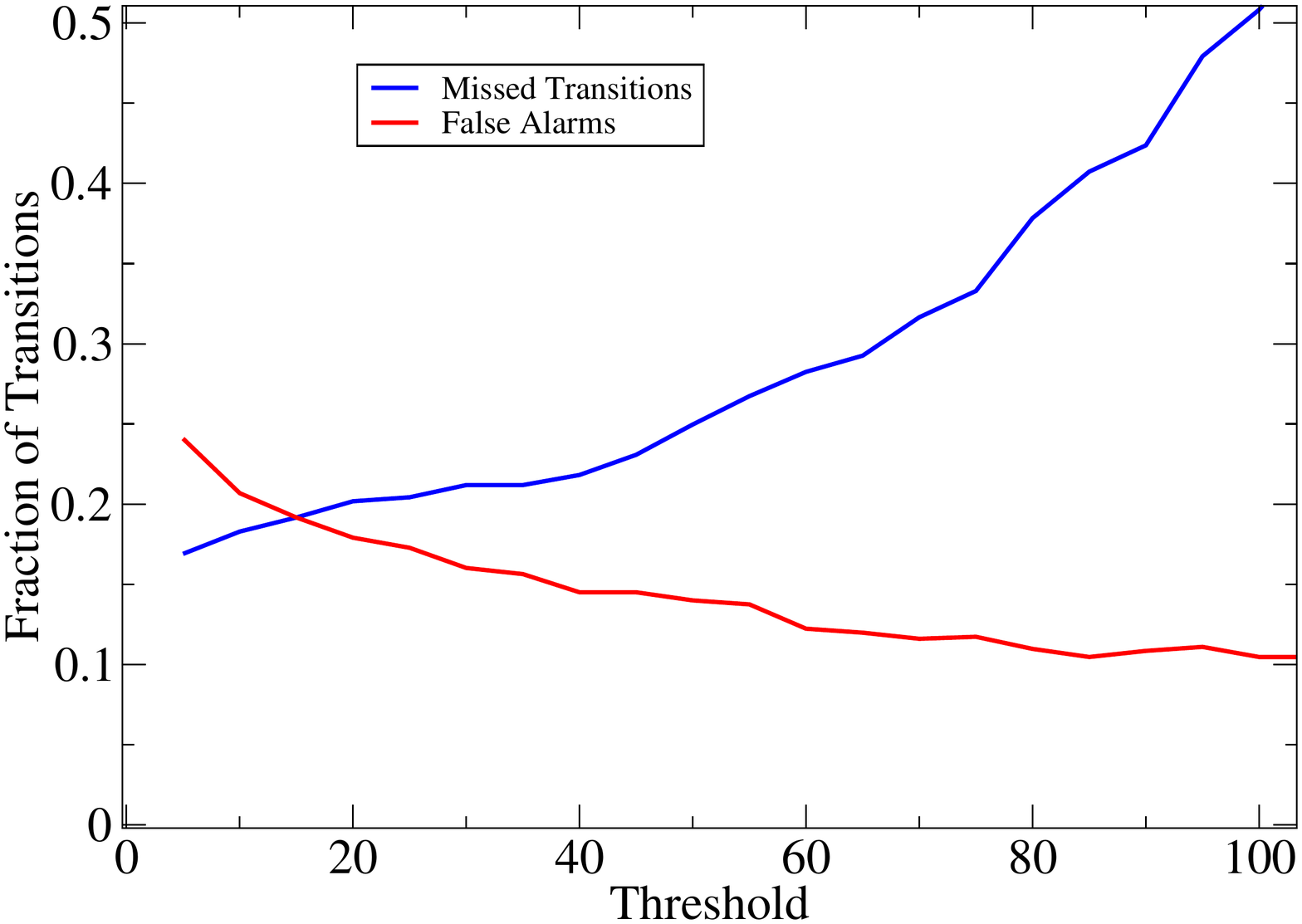}
\caption{ We can see the behaviour of the fraction of false alarms and missed transitions for different values of $A_Q$ in the Replicator Model (left panel) and the Tangled Nature (right panel). One can see how the procedure produced no false alarms in the Replicator Model which is consistent with what we expected given the Langevin nature of the model. }
\label{fig:missed_false}
\end{figure}

In Fig.(\ref{fig:missed_false}) we show these two quantities for different $A_Q$. The red curve is the fraction of missed transitions while the blu is the fraction of transitions that have produced false alarms. In the Tangled Nature Model when increasing $A_Q$ the fraction of false alarms decreases, as expected, while the fraction of missed transitions increases. The same figure for the Replicator Model shows how the procedure, although missing an increasing number of transitions, produce no false alarms at all. 

The reason for this, we believe has to do with the Langevin nature of the dynamics in the Replicator Model, i.e. deterministic dynamics + stochastic noise. Within this approach we expand the configuration vector $\vec{n}(t)$ in the $\mathbb{M}$' s eigenspace or generalized eigenspace plus noise. One gets
\begin{equation}
\vec{n}(t) = \sum_k \left( c_k(0)\text{exp}(\lambda_k t)  \vec{e}_k +\epsilon_k\right)
\end{equation}
where $c_k(0)$ are the coefficients of the expansion and $\epsilon_k$ is the noise. This dynamics is clearly dominated by those components for which $\text{Re}(\lambda_k)>0$, but this is true only if $c_k(0) \neq 0$. When a node is populated by a mutation, in our framework this corresponds to setting $c_k(t) >0$. From then on the term is suppressed if and only if the $\epsilon_k$ points in the opposite direction at all times which is highly unlikely. The same picture is less applicable to the Tangled Nature where all updates are stochastic and hence the separation in to a robust deterministic part perturbed by a weak stochastic part is problematic.
\onecolumngrid
\vspace{1cm}
\noindent\makebox[\linewidth]{\rule{\textwidth}{1pt}} 

\twocolumngrid

\begin{figure}[ht!]
\centering
\includegraphics[width=8cm,height=5.34cm]{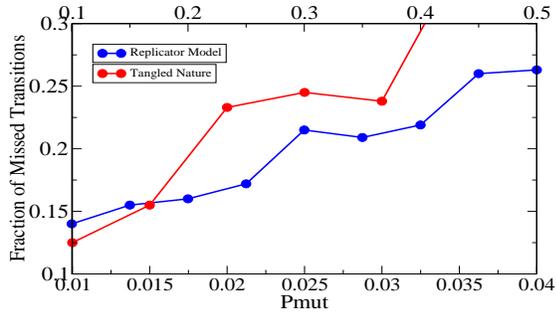}
\caption{ We present in this figure for both models the fraction of missed transitions as a function of the noise in the system. We can see how for nosier systems its harder to forecast a transition.}
\label{fig:missed_noise}
\end{figure}

The way to interpret the missed transitions is to think of the fixed points as local minima of a heterogeneous high dimensional energy landscape. The eigenspace of the mean field matrix tells where the \emph{downhill slopes} and \emph{uphill barriers} are. Although it is far more likely for the system to leave the fixed point through a downhill slope, a stochastic perturbation may be able to push the system over a barrier. This interpretation is confirmed by Fig.(\ref{fig:missed_noise}) where we can see that the fraction of missed transitions increases in both models as the degree of stochasticity is increased. 

\begin{figure}[ht!] 
\centering
\includegraphics[width=8cm,height=5.34cm]{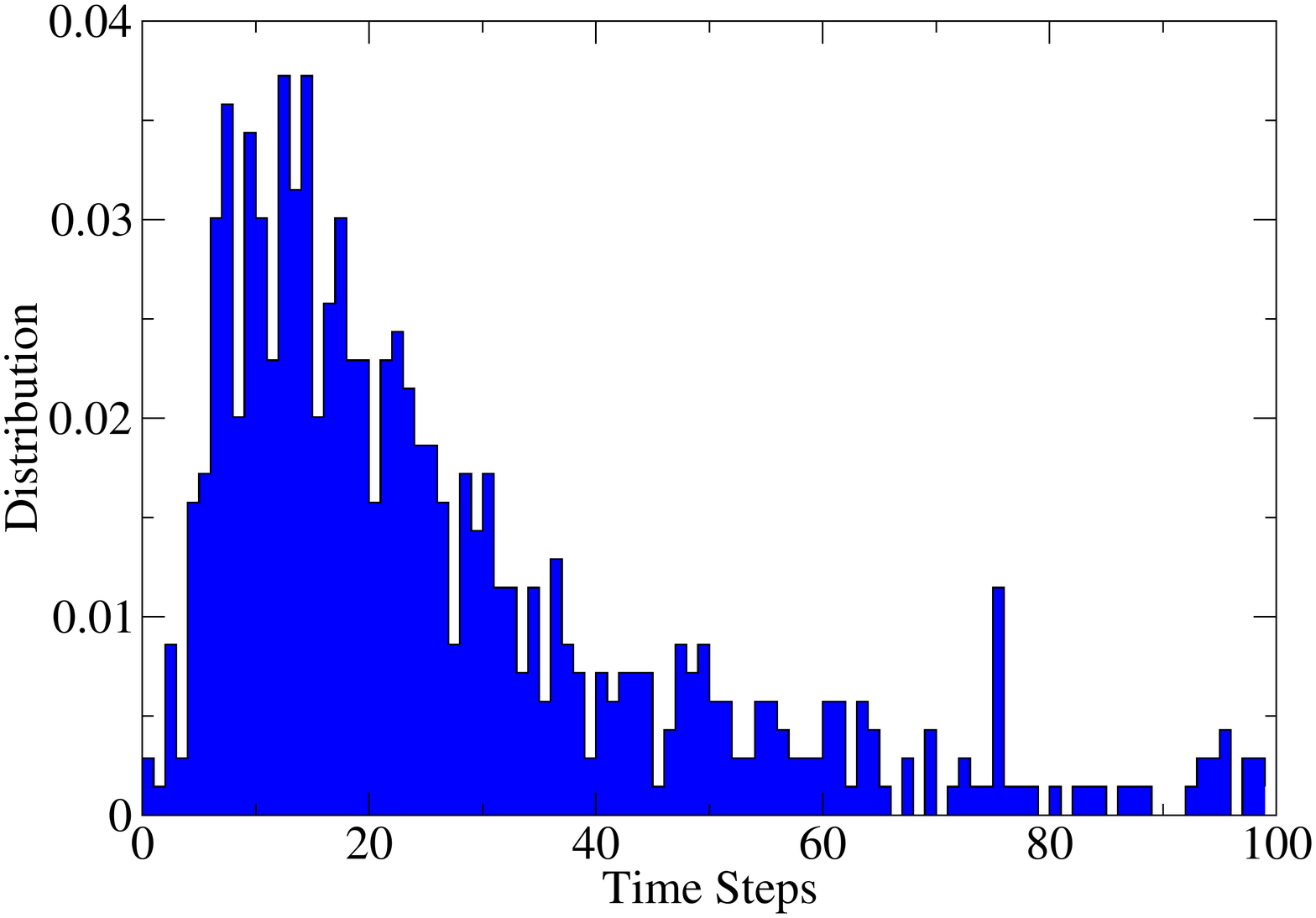}
\includegraphics[width=8cm,height=5.34cm]{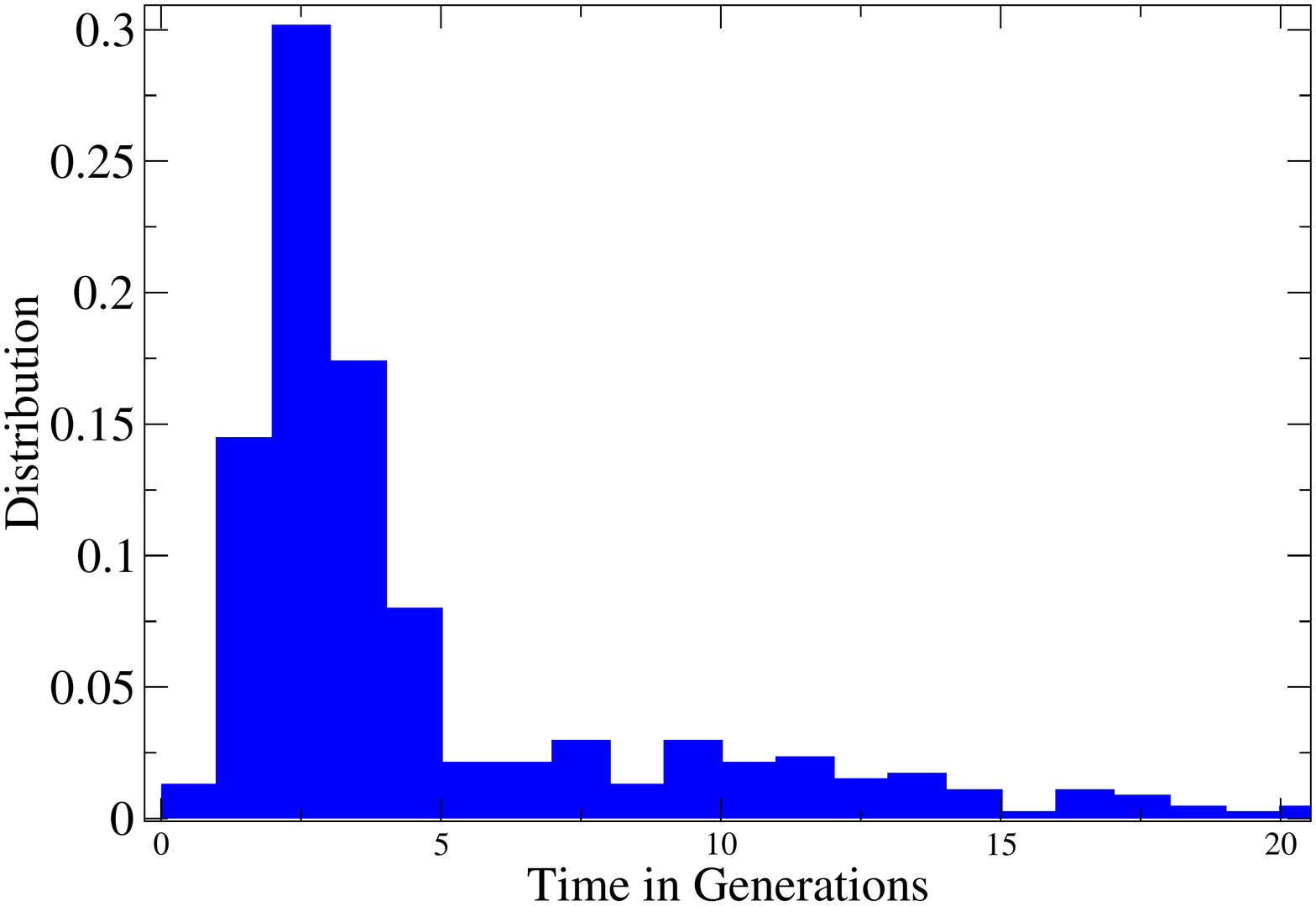}
\caption{ Distribution of the respite of the alarms for a given threshold. The left panel refers to the Replicator Model, for which $A_Q=0.01$ and the right panel to the Tangled Nature Model, for which $A_Q=20$.}
\label{fig:time}
\end{figure}

Once one fixes $A_Q$ we can check the number of time steps, $\Delta T = \|t^*-t_{cross}\|$, prior to $Q(t)$ goes above $A_Q$. In this way we can check the forecasting power of the indicator. In Fig.(\ref{fig:time}) we present the distribution of $\Delta T$ for $A_Q= 0.01$ and $A_Q = 20$ respectively for the Replicator and the Tangled Nature Model. We can see that in the Replicator Model the crossing times are tenths of time steps before the transition time. This means that the system will go through many cycles of updates before the transition occurs. In the Tangled Nature in more than $50\%$ of cases $\Delta T \in [2,5]$. As explained above when introducing the model, one generation corresponds to average number of time steps necessary to remove everyone from the system, i.e. $\frac{N(t)}{p^\text{kill}}$ individual updates. So even low values of $\Delta T$ can be considered to correspond to a strong forecasting power. 
\onecolumngrid

\vspace{.5cm}
\noindent\makebox[\linewidth]{\rule{\textwidth}{1pt}} 
\vspace{.5cm}
\twocolumngrid

\section{Incomplete Knowledge}

An obvious short coming concerning application to real situations of the forecasting procedure as described so far is that we make use of complete knowledge of the entire (both the actually realized and the "in potentia" part of) space of agents and their interactions. In this section we first consider how the lack of full knowledge of the interaction strength between agents influences our ability to detect approaching transitions. We next consider a much simpler measure than the overlap function $Q(t)$. This new measure is inspired by the analysis presented above and leading to $Q(t)$ but avoids access to information about the adjacent possible, i.e. information about agents that are not extant in the system at the tie of forecasting. Our new measure only makes use of the time evolution of directly observable quantities and can therefore in principle be applied without the need of a dynamical model of the considered system.
 
We investigate the effect of lack of complete information concerning the iterations between agents by introducing an error in the interaction matrix used for the mean field treatment. We do this in the following way 
\begin{equation}
J^{e}_\text{ij} =  J^{\text{sim}}_\text{ij} +\chi  
\end{equation}
where $\chi$ is $N(0,\sigma)$, i.e. a normally distributed random variable, of mean $0$ and variance $\sigma$. We then repeat the exact same procedure outlined in the previous section but using $\mathbb{J}^\text{e}$ in the calculations. 
\onecolumngrid

\begin{figure}[ht!] 
\centering
\includegraphics[width=8cm,height=5.34cm]{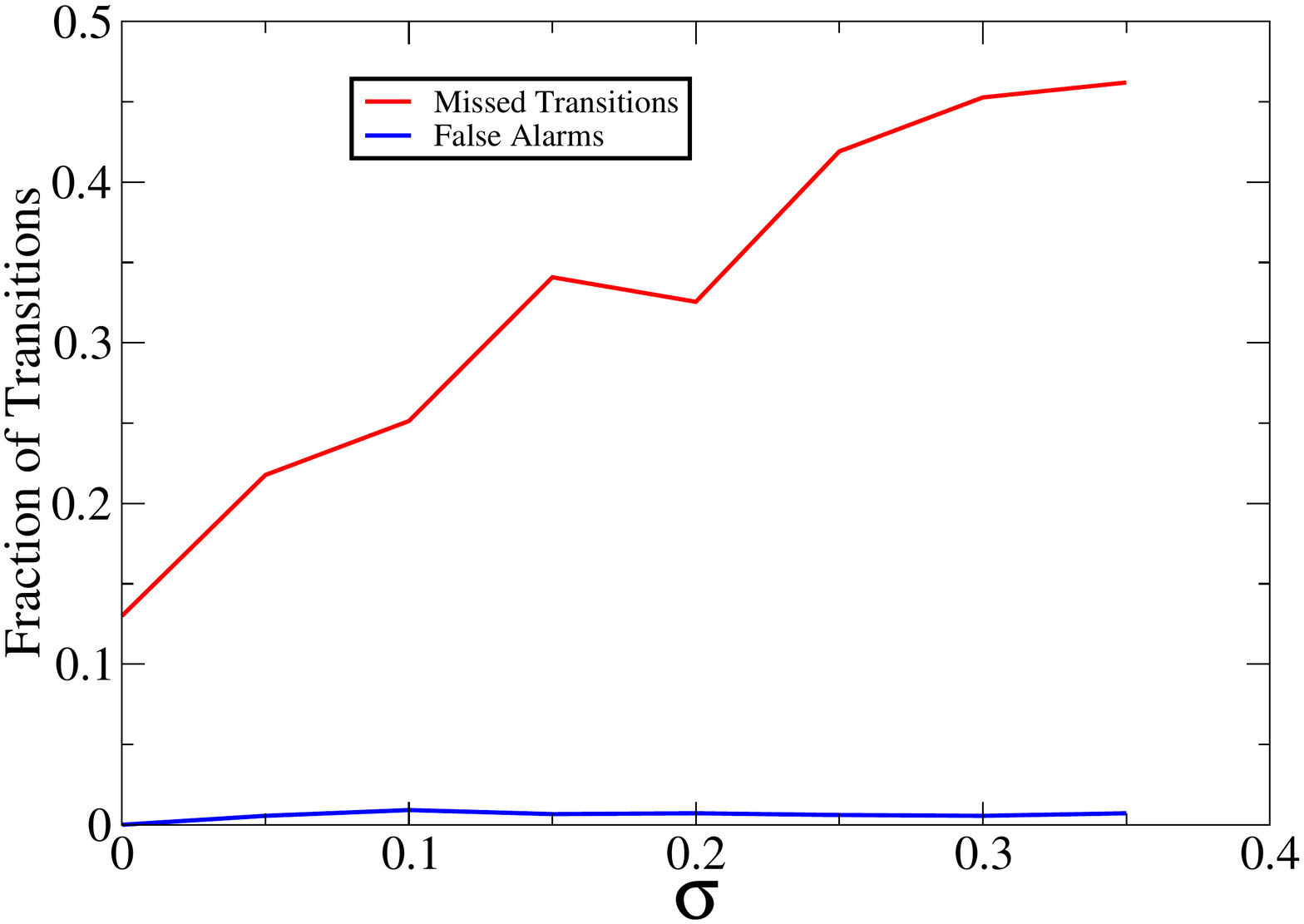}
\includegraphics[width=8cm,height=5.34cm]{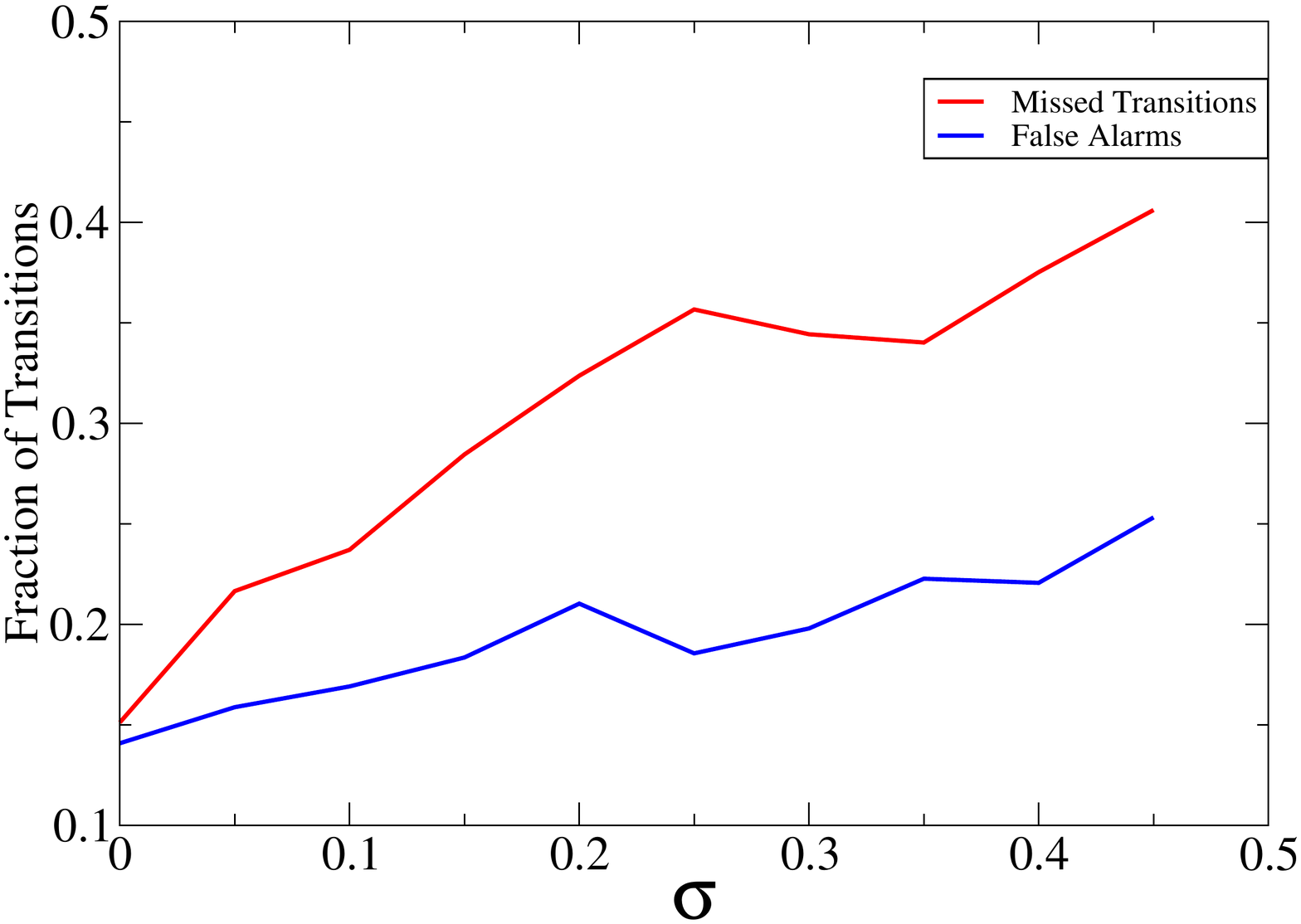}
\caption{In this figure we show the fraction of the transitions we are not able to forecast and the fractions of false positive, in function of the $\sigma$ of the distribution of the random error in the interactions. Once again we have used $A_Q = 30$ for the Tangled Nature (right panel) and $A_Q= 0.01$ of the Replicator Model (left panel).}
\label{fig:error}
\end{figure}

\begin{figure}[!h] 
\includegraphics[width=8cm,height=5.34cm]{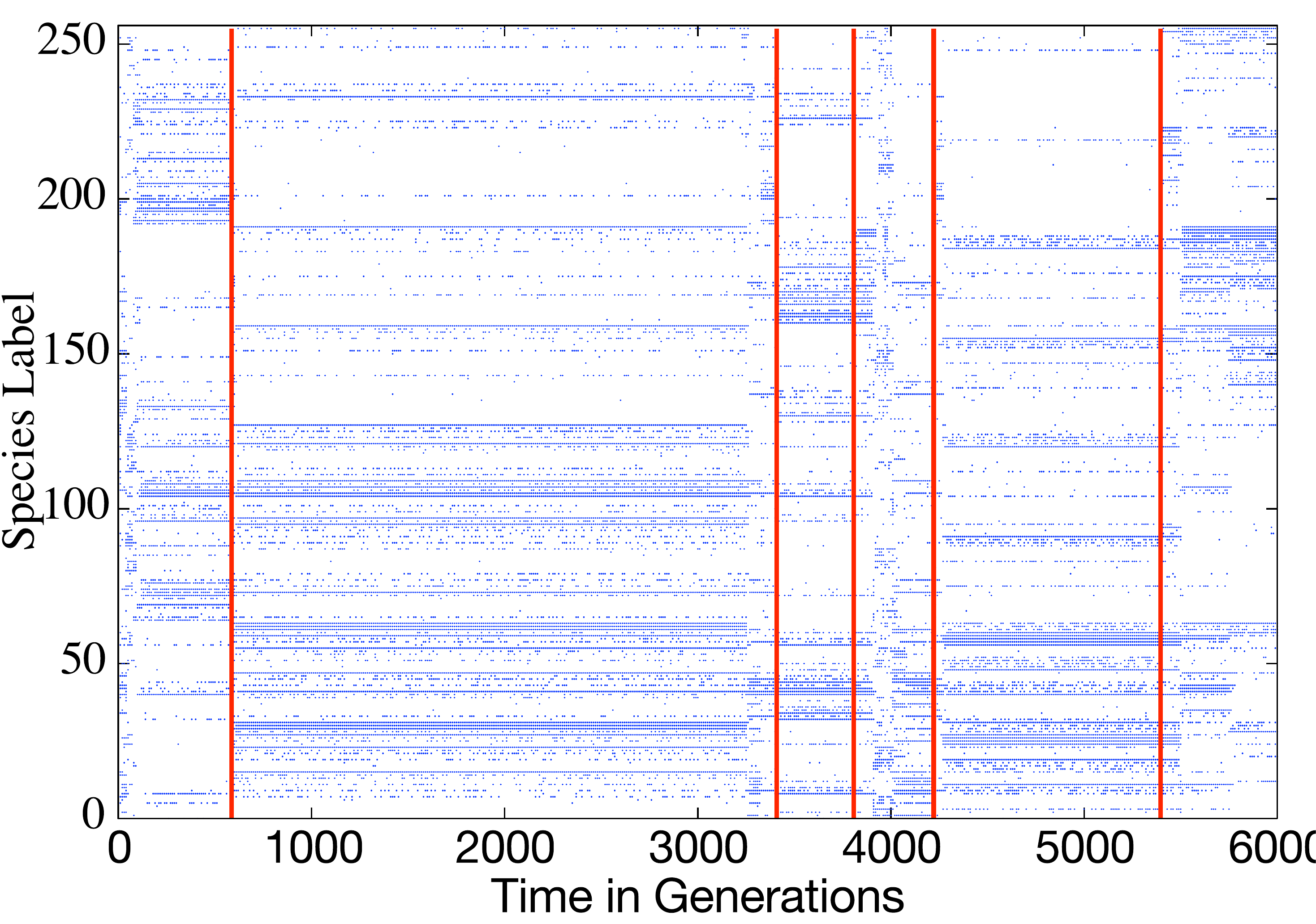}
\includegraphics[width=8cm,height=5.34cm]{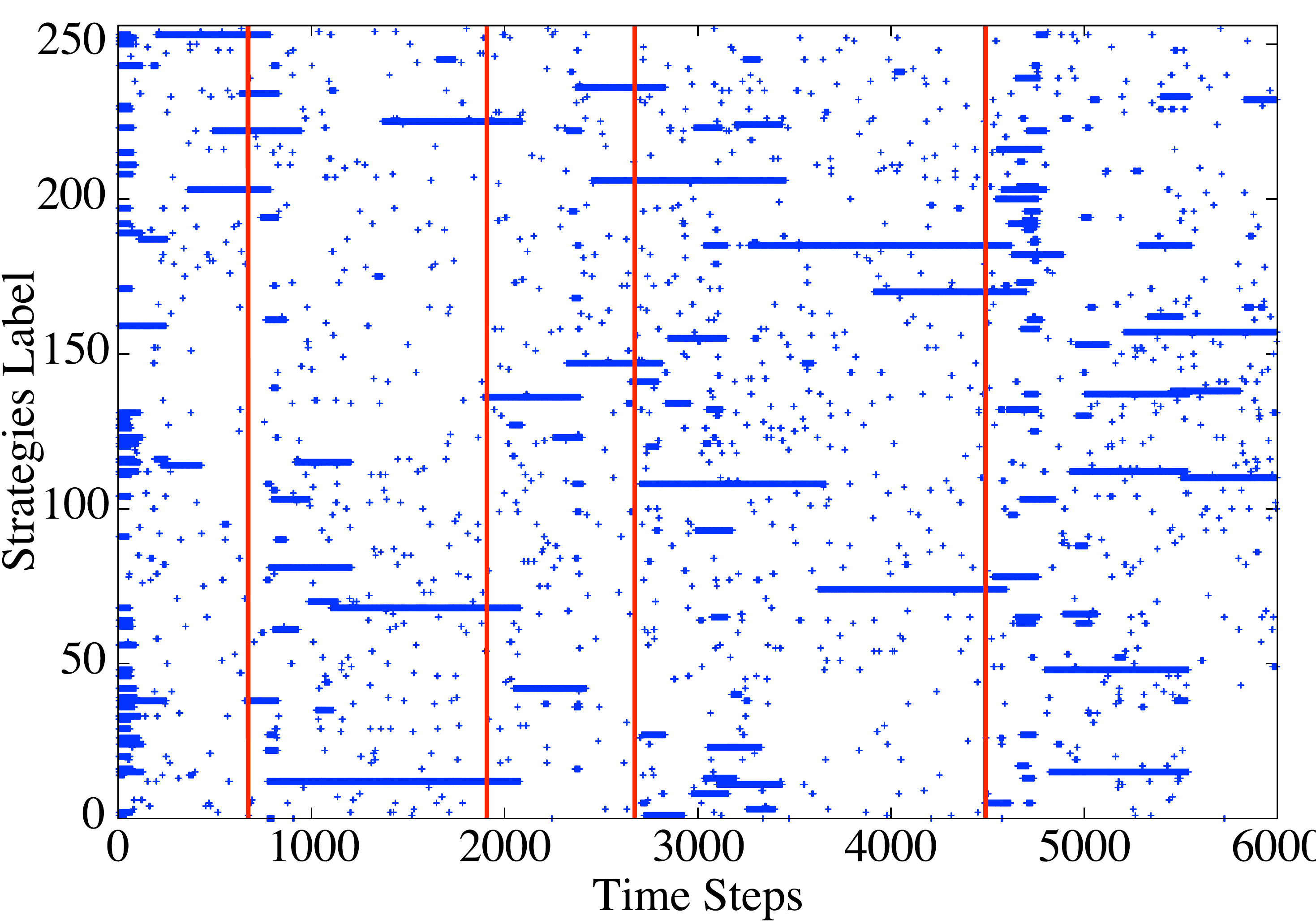}\\
\includegraphics[width=8cm,height=5.34cm]{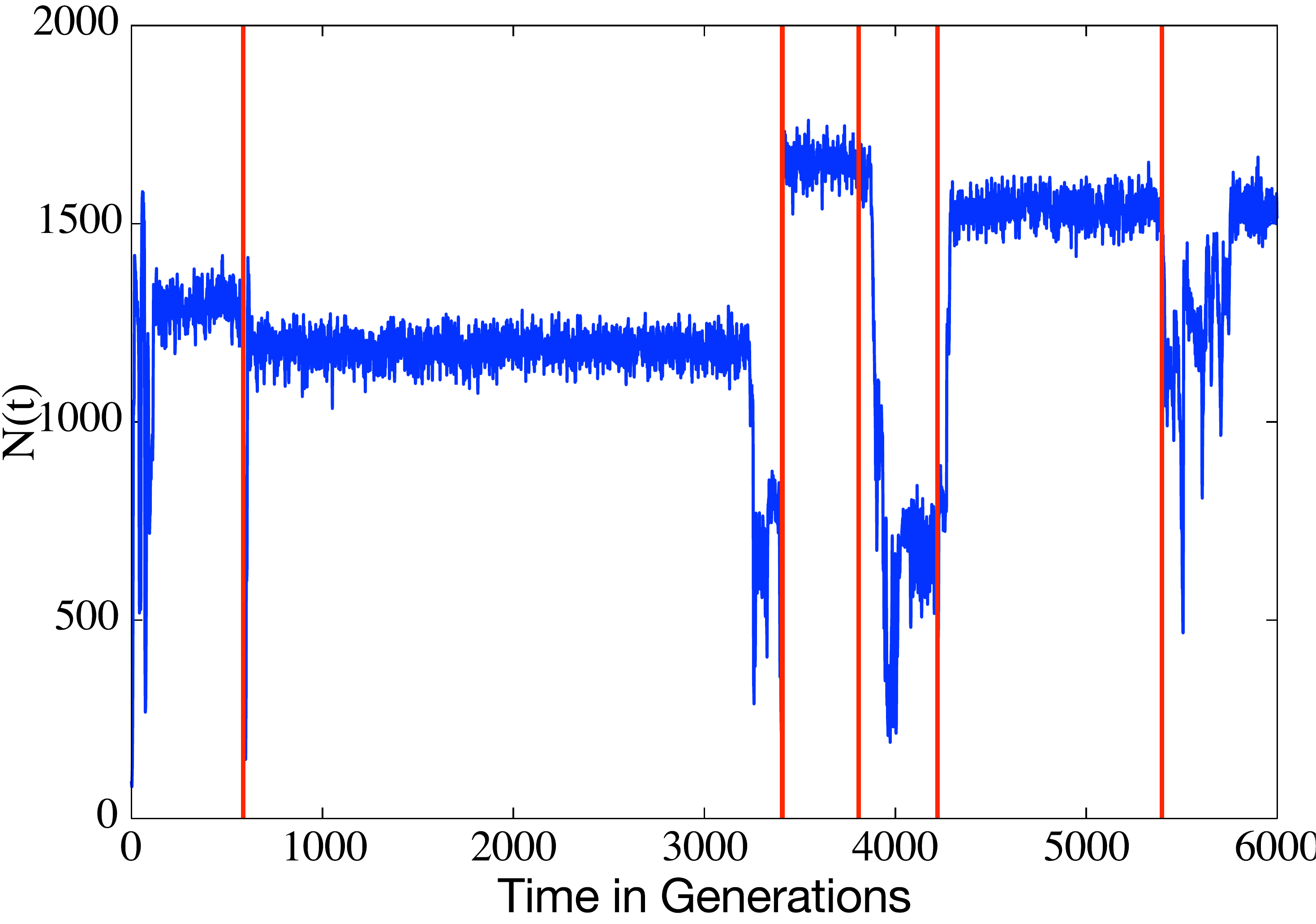}
\includegraphics[width=8cm,height=5.34cm]{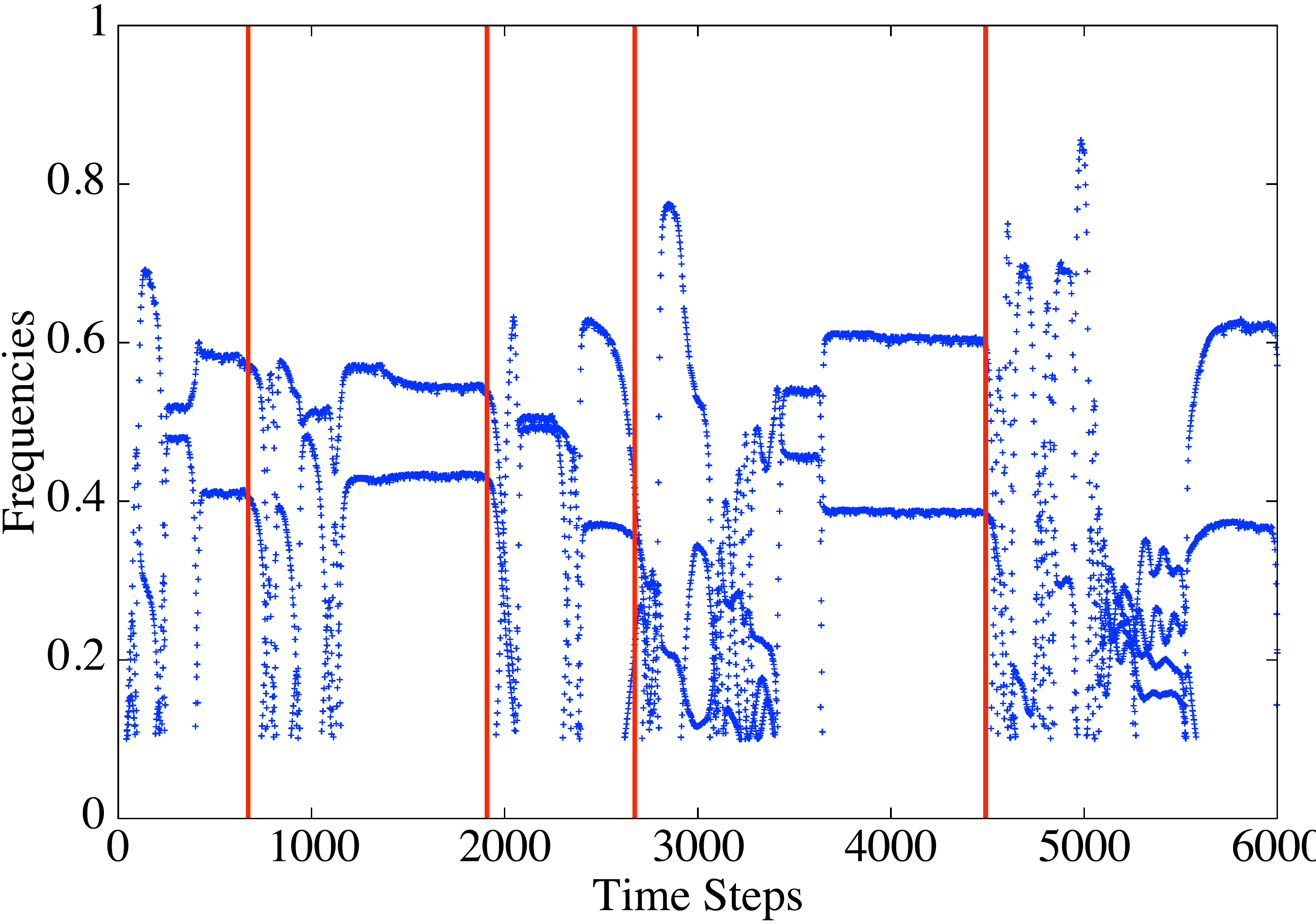}
\caption{
Top left and bottom left respectively occupation plot and total numbers of individual $\sum_j n_j(t) = N(t)$ in the Tangled Nature Model. The vertical red lines represent the alarm times. In the top and bottom right we compare the behavior of the occupation plot and the frequencies of the most occupied strategies (blue curves) in the Replicator model with the alarms given by our new procedure . One can clearly see how after every alarm the system changes its configuration.}
\label{fig:alarms}
\end{figure}
\noindent\makebox[\linewidth]{\rule{\textwidth}{1pt}} 
\vspace{.5cm}
\twocolumngrid

In Fig.(\ref{fig:error}) we present the fractions of transitions we are not able to forecast and the fractions of false alarms we generate as function of the variance $\sigma$, i.e. as function of how much the interaction matrix used for the stability analysis differs from the correct set of interactions. For the Tangled Nature (see the right panel) we can notice that for $\sigma<0.2$ we are still able to forecast around $70\%$ of the transitions and we generate less than $20\%$ of false alarms. This is an encouraging result since a 
$\sigma = 0.2$ is clearly a significant error given that $J_\text{ij} \in \left(-1,1 \right)$. A very similar result holds for the Replicator Model. 

We now discuss a forecasting procedure that doesn't need any knowledge about "in potentia" agents. We only need to focus on the highly occupied nodes present in the system. We only know what we see without making any use of the non active part of the interaction network, nor of the poorly occupied nodes. 

By applying the LSA to the occupied network we can check that, during a stable phase, the configuration corresponds to a situation where the spectrum of the $\mathbb{M}$ consists of eigenvalues that all have negative real parts. As the system evolves new mutants appear. As an indicator of approaching transitions we track the growths of the occupancy of these new agents, if their occupancy exceeds a certain threshold we check the spectrum of the updated $\mathbb{M}$, in which the new agents are included. In case the spectrum now includes positive eigenvalue we take this as an indicator of, an approaching transition out of the present metastable configuration. This will be our new alarm.
 
In Fig.(\ref{fig:alarms}) we show the results of an application of this new procedure. In both panels the red vertical lines indicate the times of appearance of a species able to change the stability of the system. We can qualitatively see from the figure that just after the alarms the system actually undergoes a transition.

In the left panel of the Fig.(\ref{fig:alarms}) the blue curves represent the frequencies of the most occupied strategies in the Replicator model. We can see how right after the red lines, the alarm times, a new strategy starts gaining frequency and eventually puts an end to the stable configuration. 

In the right panel we show the total number of individuals present in the system $N(t) =\sum_j n_j(t)$. A transition to a new metastable configuration is associated with a sudden change of this quantity. We notice that after each alarm $N(t)$ exhibit a significant change. Preliminary analysis indicates that this procedure is able to forecast transitions with an accuracy similar $Q(t)$ indicator. Further investigation of the efficiency and reliability of using the grows rate of new comers as indicators of approaching transitions is underway. Obviously this can make our procedure more readily applicable to real systems since we would then only need directly observable information.

\section{Summary and Conclusion} 
We have describe a new procedure for forecasting transitions in high dimensional systems with stochastic dynamics. Our method is of relevance to systems where the macroscopic dynamics at the systemic level is not adequately captured by a well defined set of essentially deterministic collective variables (e.g. as handled by Langevin equations). Hence we are dealing with situations that are not captured by the application of bifurcation theory such as considered by Scheffer and collaborators ~\cite{scheffer2009critical,scheffer2012anticipating,scheffer2009early}. We have in mind complex systems in which the dynamics involves some evolutionary aspects, in particular situations where the dynamics generates new degrees of freedom. E.g. biological evolution, or economical and financial systems, where new agents (organisms, strategies or companies, say) are produced as an intrinsic part of the dynamics. We have demonstrated by use of two models of varying degree of stochasticity (the Tangled Nature Model and the stochastic Replicator Model)
that a combination of analytic linear stability analysis and simulation allows one to construct a signal (overlap with unstable directions) which can be used to forecast a very high percentage of all transitions. 

The weakness of the procedure is that for real situations of interest (e.g. an ecosystem or a financial market) one may obviously not posses complete information. One will typically not have access to all the information about the interaction amongst the agents. This turns out to be less of a problem, since we can show that even with a 10\% inaccuracy in interaction strengths, we are still able to forecast a substantial percentage of transitions. Another short coming is that in real situations it can also be very difficult to know the nature of the new agents that may arrive as the system evolve. Our full mathematical procedure suggests a way to overcome this problem. Namely, the eigenvector analysis showed that transitions are often accompanied by the arrival of new agents, which exhibit a rapid growth in their relative systemic weight. We found that simply monitoring the rapidly growing new agents can enable prediction of major systemic upheavals. I.e. approaching transitions might not be apparent by focusing on the systemic heavyweights, but rather one should keep a keen eye on the tiny components to monitor whether they suddenly start to flourish. This can often be the signal of upcoming systemic changes. Our next step will be to test these findings on real data streams including high frequency financial time series.

\vspace{1cm}
\textit{Acknowledgments} - This work was supported by the European project CONGAS (Grant FP7-ICT-2011-8-317672). We are grateful for computer time on Imperial College's High Performance Cluster.

\end{document}